\documentclass[11pt]{article}
\pdfoutput=1

\usepackage[verbose=true,letterpaper]{geometry}
\AtBeginDocument{
  \newgeometry{
    textheight=9in,
    textwidth=6in,
    top=1in,
    headheight=14pt,
    headsep=25pt,
    footskip=30pt
  }
}

\usepackage{graphicx} 
\graphicspath{ {images/} }

\usepackage[square,numbers,sort&compress]{natbib}
\usepackage{hyperref}
\hypersetup{colorlinks,allcolors=black}

\usepackage[linesnumbered,ruled,vlined]{algorithm2e}
\usepackage[noend]{algpseudocode}
\usepackage{subcaption} 
\usepackage{booktabs} 
\usepackage{amsfonts}
\usepackage{amssymb}
\usepackage{amsmath}
\usepackage{bm}
\usepackage{url}
\usepackage[inline]{enumitem}

\title{METASET: Exploring Shape and Property Spaces for Data-Driven Metamaterials Design}

\author{Yu-Chin Chan \quad Faez Ahmed \quad Liwei Wang \quad Wei Chen\thanks{Address all correspondence to this author. Email: weichen@northwestern.edu}\\ \\
	Dept. of Mechanical Engineering\\
	Northwestern University\\
	Evanston, Illinois, USA
}

\date{\vspace{-5ex}}

\newcommand{\eg}{{\em e.g.}}
\newcommand{\ie}{{\em i.e.}}

\newcommand{\etal}{{\em et~al.}}

\begin{document}
\maketitle    

\begin{abstract}
{\it Data-driven design of mechanical metamaterials is an increasingly popular method to combat costly physical simulations and immense, often intractable, geometrical design spaces. Using a precomputed dataset of unit cells, a multiscale structure can be quickly filled via combinatorial search algorithms, and machine learning models can be trained to accelerate the process. However, the dependence on data induces a unique challenge: An imbalanced dataset containing more of certain shapes or physical properties can be detrimental to the efficacy of data-driven approaches. In answer, we posit that a smaller yet diverse set of unit cells leads to scalable search and unbiased learning. To select such subsets, we propose METASET, a methodology that \begin{enumerate*} \item[1)] uses similarity metrics and positive semi-definite kernels to jointly measure the closeness of unit cells in both shape and property spaces, and \item[2)] incorporates Determinantal Point Processes for efficient subset selection.
\end{enumerate*} Moreover, METASET allows the trade-off between shape and property diversity so that subsets can be tuned for various applications. Through the design of 2D metamaterials with target displacement profiles, we demonstrate that smaller, diverse subsets can indeed improve the search process as well as structural performance. By eliminating inherent overlaps in a dataset of 3D unit cells created with symmetry rules, we also illustrate that our flexible method can distill unique subsets regardless of the metric employed. Our diverse subsets are provided publicly for use by any designer.\footnote{\url{https://github.com/lychan110/metaset}}}
\end{abstract}

\section{Introduction}\label{sec:intro}
Metamaterials are drawing increased attention for their ability to achieve a variety of non-intuitive properties that stem from their intentionally hierarchical structures~\cite{Schumacher2015}. While they traditionally consist of one unit cell that is repeated everywhere, multiple unit cells can also be assembled to create \textit{aperiodic} mechanical metamaterials with, \eg, spatially-varying or functionally-gradient properties~\cite{Schumacher2015,Maskery2018tpms}. Over the past few years, conventional computational methods have been adapted to design these complex structures, including topology optimization (TO) of the microscale unit cells within a fixed macroscale structure~\cite{Choi2019preallocate,Vogiatzis2018conformal}, and hierarchical and concurrent multiscale TO that design both the macrostructure and a pre-specified number of unique unit cells~\cite{Deng2017concurrent,Du2018connectivity,Liu2019subdomain}. However, as the desire to attain even more intricate behaviors grows, so too does the complexity of the design process, which must account for the expensive physical simulations and, in aperiodic structures, the vast combinatorial design space and disconnected neighboring unit cells~\cite{Schumacher2015,Coulais2016combinatorial}.

\begin{figure}[t]
    \centering
    \includegraphics[width=0.7\columnwidth]{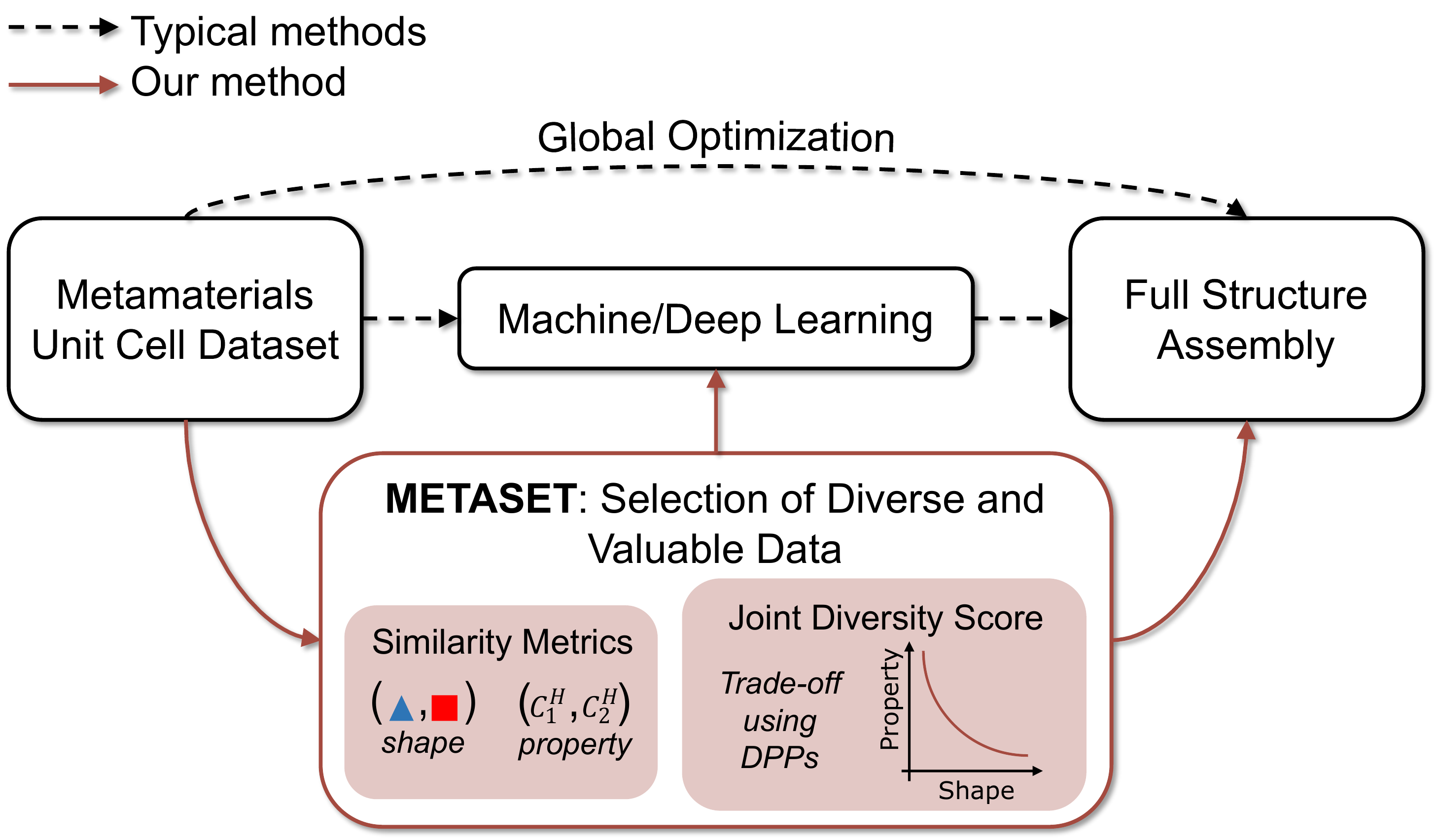}
    \caption{A high-level overview of data-driven metamaterials design, and how our proposed method, METASET, fits in. As an example, we show $C^H$, the homogenized elastic tensor, as the unit cell properties.}
    \label{fig:flow}
\end{figure}

Capitalizing on advances in computing power, data-driven metamaterials design can be a more efficient and therefore enticing solution to those challenges. Its success hinges on precomputed unit cell libraries or datasets, which can avoid costly on-the-fly physical simulations and multiscale TO in huge design spaces, as well as provide candidate unit cells that are better connected to their neighbors. Fig.~\ref{fig:flow} shows an overview of two common approaches in data-driven design: global optimization methods, and machine learning (ML) based methods. In the first case, combinatorial optimization algorithms can be used to directly search for the set of unit cells that realize a target macroscale behavior while minimizing or constraining the boundary mismatch between neighboring cells ~\cite{Schumacher2015,Coulais2016combinatorial,Zhu2017twoscale}. From another perspective, data-driven methods can use the dataset to train ML models that further accelerate design. For example, they have been used to rapidly predict homogenized physical properties as part of the optimization loop~\cite{Wang2020smo,Bostanabad2019gagp,White2019neural,Chen2015coarse}. Additionally, deep generative models inspired by the computer vision field can learn embedded geometric descriptors that act as reduced dimensional design variables, and construct new designs, \eg{}, optical 2D metamaterials~\cite{Ma2019vae,Liu2018nano}, almost instantaneously. Accelerated by data-driven techniques, challenging designs such as spatially-varying displacement profiles and nonlinear behavior that are prohibitively expensive via conventional methods are now tangible.

The efficacy of data-driven methods, however, relies highly on the size and coverage of the datasets. The search space of global optimization methods can quickly explode when the number of unit cells increases. Meanwhile, imbalanced datasets with skewed data distributions can reduce the chance of meeting certain property or compatibility requirements, and hobble the performance of ML models since they may not learn a less frequent property or shape as well~\cite{Haixiang2017imbalance}. Therefore, due to the importance of the data on downstream tasks, in this work we focus on the first step of data-driven design: dataset selection.

In existing literature, metamaterial datasets are often built using heuristics or the designer's intuition, with the assumption that the unit cells will offer sufficient coverage for the desired application. Many employ TO to inversely design unit cells that meet pre-specified target properties~\cite{Schumacher2015,Zhu2017twoscale,Wang2020smo}, and some expand the dataset by morphing the shapes~\cite{Schumacher2015,Wang2020smo} or randomly flipping pixels or voxels~\cite{Zhu2017twoscale}. Alternatively, Panetta~\etal{} developed graph-based rules to create truss-like unit cells~\cite{Panetta2015}. Although these are more feasible than enumerating over all possibilities, bias toward particular properties or shapes can be unintentionally introduced, deteriorating the performance of the design algorithm or the design itself.

Moreover, the point at which to stop generating new unit cells has thus far been heuristic with the same goal in mind: to cover a broad property space. The range of this space is sometimes restricted for specific applications~\cite{Choi2019preallocate}, or strict symmetry and manufacturability constraints are implemented to limit the possible shapes~\cite{Panetta2015}. More often, the property space is allowed to grow at will, \eg{}, TO and shape perturbation are repeated until the change in the density of the property space is less than a given tolerance~\cite{Zhu2017twoscale,Wang2020smo}. While efficient, all of the works to date have only considered coverage in the property space alone, which can produce similar shapes or overlook those that might benefit the design with regards to boundary connectivity. In contrast, our work explores coverage in both property and shape spaces.

Improving imbalance arising from data with multiple classes has been extensively researched in computer science. The most relevant to our application are the data preprocessing strategies such as undersampling to remove data from majority classes, oversampling to replicate data from minority classes, or combinations thereof~\cite{Haixiang2017imbalance}. However, the former can accidentally remove samples with important features, \ie{}, decrease the diversity, and the latter can lead to model overfitting and increased training overhead~\cite{Branco2016imbalance2}. Nor are they made to consider the diversity of data with features that have drastically different representations, like shape and property. The issue of downsampling a metamaterial database was addressed by Chen~\etal{}~\cite{Chen2015coarse}, who compressed the size of their database by selecting the samples that are farthest from each other with respect to properties (not shape), allowing them to more efficiently fit a property prediction model. As far as we know, there is currently no method to assess or select a diverse set of unit cells that can simultaneously cover the shape and property spaces. 

Despite the dearth in the metamaterials field, measuring and ranking items based on their quality as well as their contribution to the diversity of a whole set or subset is an ongoing research area. In computer science, for example, recommender systems rank diverse items such as online products to match users' preferences. These are based on the concept of diminishing marginal utility~\cite{Coombs1977diminishing}, wherein lower ranking items bestow less additional value onto the users. In design, too, researchers have developed methods to help designers sift through large sets of ideas by ranking them. In particular, to balance diversity against quality of designs, Ahmed~\etal{} introduced the idea of clustering items into groups for subset selection~\cite{ahmed2016discovering} by employing submodular functions that follow the property of diminishing marginal utility. Additionally, Ahmed~\etal{}~\cite{Ahmed2017ranking} showed the application of Determinantal Point Processes (DPPs)~\cite{kulesza2012determinantal}, which model the likelihood of selecting a subset of diverse items as the determinant of a kernel matrix, to the diverse ranking task. The latter, in particular, are elegant probabilistic models that capture the trade-off between competing ideas like quality and diversity. While the goal of maximizing the determinant is similar to the optimality criterion used in generating D-optimal designs~\cite{de1995d} in design of experiments, DPPs are not restricted to linear kernels, and have advantages in that calculating marginals, computing certain conditional probabilities and sampling can all be done in polynomial time. This paper shows that DPPs can also be used for coverage in multiple spaces defined over the shapes and properties of unit cells.

\textbf{Our contributions:}
We propose METASET, an automated methodology that simultaneously considers the diversity of shape and property to select subsets of unit cells from existing datasets. By doing so, we can achieve scalable data-driven design of metamaterials using smaller yet diverse subsets and eliminate bias in imbalanced datasets to improve any downstream task in the data-driven framework. As a part of METASET, we introduce similarity metrics to efficiently assess the diversity of the shapes and properties of 2D and 3D metamaterials. We also propose that a weighted sum of Determinantal Point Process (DPP) kernels based on the shape and property similarities can measure and allow the maximization of the joint diversity of both spaces. 
For the first time in data-driven metamaterials design~---~to our knowledge~---~we reveal through 2D case studies that diverse subsets can expedite and even enhance the design performance and connectivity of aperiodic metamaterials. 
Finally, applying METASET to 3D unit cells, we identify diverse families of isosurface unit cells and discover that these extend beyond the ones commonly considered in the design of functionally-graded structures~\cite{Li2019tpms,Maskery2018tpms}.

The components of our methodology are detailed in Sec.~\ref{sec:methods}.
In our 2D case studies (Sec.~\ref{sec:2dcase}), we explore the effects of diversity and subset size on 2D metamaterial designs with non-intuitive target displacement profiles. In a 3D example (Sec.~\ref{sec:3dcase}), we compare the impact of different shape similarity metrics on diverse unit cell families and demonstrate that METASET can diversify datasets regardless of the chosen metric.

\section{METASET: Assessing and Optimizing Diversity}\label{sec:methods}
The inner workings of METASET consist of three main steps:
\begin{enumerate*}
    \item[1)] Defining similarity metrics for metamaterials that quantify the difference between pairs of 2D or 3D shapes and mechanical properties (Sec.~\ref{sec:similarity});
    \item[2)] Using a DPP-based submodular objective function to measure the joint coverage of a set of unit cells in shape and property spaces via pairwise similarity kernel matrices (Sec.~\ref{sec:dpp});
    \item[3)] Maximizing the joint diversity with an efficient greedy algorithm while allowing trade-off in the two spaces to be tuned to suit the desired application (Sec.~\ref{sec:ranking}).
\end{enumerate*}
In this section, we describe these components and summarize the methodology with Algorithm \ref{alg:alg_metaset}.

\subsection{Similarity Metrics for Metamaterials}\label{sec:similarity}
A diverse metamaterial dataset should ideally contain unit cells that are sufficiently different, \ie{}, dissimilar, such that they cover the shape and property spaces. To measure the diversity of a set, then, the similarities between the shapes and properties of unit cells first need to be quantified. We do so by defining metrics independently in each space, based on the observation that a set of unit cells dissimilar in shape space is not necessarily also dissimilar in property space, and vice versa. 
This can be illustrated by a simple example. Say we wish to distill diverse values from $x$ and $y$, which we assume to be sets of integers: $x = \{0, 1, 2, 4, 5\}$ and $y = \{0, 2, 10, 20, 10\}$. We assume that $y =  x*k$, where $k = \{3, 2, 5, 5, 2\}$ is a transformation function. If we were to select three diverse values of $x$, \ie{}, the values that most cover its space, we would select $\{0, 2, 5\}$. For $y$, however, we would choose $\{0, 10, 20\}$ rather than $\{0, 10, 10\}$, the ones corresponding to the diverse $x$ values. Hence, though some relationship between two spaces may exist, \eg{}, an intrinsic function between shape and property, there is a need to model their coverage separately. This observation is validated in our later design experiments (Sec.~\ref{sec:2ddpp}), where the correlation coefficient between shape and property coverage shows that no link exists between the two.

\subsubsection{Property Similarity} 
Since mechanical properties are generally scalar values that can be expressed as a vector, \eg{}, by flattening the elastic tensor, we can use any similarity metric between vectors. In this work, we use the Euclidean distance. We note that the properties do not need to be the tensor components; rather, they can be other values of interest such as elastic or shear moduli, or Poisson's ratios. Neither do they need to be limited to scalar mechanical properties. For instance, dynamic acoustic dispersion curves or bandgaps could be considered if the pairwise similarity can be quantified.

\subsubsection{Shape Similarity}
Shape similarity metrics are key in many computer vision and graphics applications, \eg, facial recognition and object retrieval from databases. In these methods, the shapes are usually first represented by structural descriptors extracted from individual shapes~\cite{Bustos2005featurebased}, or by embedded features learned via data-driven methods such as clustering or deep learning~\cite{Rostami2018embedded,achlioptas18ae}. The distances between features can then be measured in Euclidean~\cite{Bustos2005featurebased} or Riemmanian space~\cite{Sharon2006conformal2d,Su2015conformal3d}. Since Riemannian metrics are based on geodesic distances, they are suitable if one needs invariance to deformation, \ie{}, if one considers a shape to be the same after bending.

For metamaterials, however, we must rule out deformation and rotation invariant metrics since any transformation of a unit cell impacts its properties. Additionally, we seek techniques that are efficient but still able to discriminate fine details and form positive semi-definite similarity matrices for the next step involving DPPs. Thus, we introduce the following Euclidean metrics based on structural features: a descriptor-based distance for 2D, and two point cloud-based metrics for 3D, namely, the Hausdorff distance and embedded cosine similarity utilizing deep learning. While we elected for separate metrics in 2D and 3D by bearing in mind their respective computational efficiencies, shape analysis is a wide and ever-growing topic of research in computer science; many other metrics are available. As we later show in Sec.~\ref{sec:3ddpp}, METASET selects diverse subsets regardless of the metric used, as long as the requirements for DPPs are met.

\vspace{6pt}
\textbf{2D Descriptor-Based Euclidean Distance: }\label{sec:2dsimilarity}
For 2D unit cells, which are typically binary images resulting from TO, we propose using a descriptor-based approach by first extracting division-point-based descriptors~\cite{vamvakas2010handwritten} to reduce the images into vectors that capture salient features at different levels of granularity. This has been applied to the field of optical character recognition~\cite{das2012statistical,sarkhel2017multi}. The binary image of a unit cell is recursively divided into sub-regions that contain an equal number of solid pixels. The coordinates of all division points, \ie{}, points at the intersection of two division lines between each sub-region, are then obtained as descriptors of the unit cell. This process is repeated until the desired level of detail is captured, constructing a \textit{k}-d tree of the distribution of solid materials. In our 2D case study (Sec.~\ref{sec:2dcase}), we obtain a sufficient amount of detail by performing the division seven times for each unit cell, resulting in 62 division points that constitute a 124-dimensional shape descriptor. 

Using the above method, we can represent each 2D unit cell as a vector, then use the Euclidean norm to find the distance between any pair. However, the input for a DPP is a positive semi-definite similarity matrix, $L$, so we transform the distance to a similarity metric through a radial basis function kernel with unit bandwidth, \ie, $L_{i,j}=\exp(-0.5~d(i,j)^2)$, where $d(i,j)$ is the distance between $i$-th and $j$-th unit cells. In practice, the choice of an appropriate transformation is equivalent to choosing the right distance metric between items. Our empirical study on other common transformations showed that different choices mainly affect the distribution of similarity values but do not significantly affect the final outcome or the key findings of our work.

\vspace{6pt}
\textbf{3D Hausdorff Distance: }\label{sec:3dsim_hausd}
As for 3D unit cells, mesh formats such as STL are commonly used so that the metamaterials can be manufactured through additive manufacturing. However, since performing analysis on 3D shapes is undoubtedly more computationally intense due to the curse of dimensionality, we suggest representing each unit cell as points on the surface of the original mesh, \ie{}, point clouds, which are more efficient for extracting and processing 3D features~\cite{Kobbelt2004pointcloud}. This extra conversion can take little computation with well-established sampling methods, \eg{}, randomly sampling the surface of a mesh with the probability of choosing a point weighted by the area of the triangular faces.

We then use a distance metric commonly utilized to measure the distance between sets of points, the Hausdorff distance. In essence, it computes the difference between two clouds as the maximum of the nearest neighbor distances of each point. This is expressed as \cite{Huttenlocher1993hausdorff}:
\begin{equation}\label{eq:hausdorff}
    h(A,B) = \max_{a \in A}{\big[\min_{b \in B}{\lVert \cdot \rVert}\big]},
\end{equation}
where $a$ is a point within cloud $A$ and $b$ is a point in the second cloud $B$. The notation $\lVert \cdot \rVert$ indicates that any distance can be used; for example, we can use the Euclidean norm or the cosine distance between two points. In our implementation, we computed the nearest neighbor Euclidean norms using a GPU-enabled code by Fan \etal{}~\cite{Fan2017nndist}. Then, to obtain a symmetric distance, we take the maximum as follows:
\begin{equation}\label{eq:hausdorff2}
    d_\text{H}(A,B) = d_\text{H}(B,A)= \max{\big[h(A,B), h(B,A)\big]}.
\end{equation}
Finally, we convert the pairwise distances into a DPP similarity kernel, $L$, using the following transformation: $L_{ij} = \frac{1}{1+d(i,j)}$.

\vspace{6pt}
\textbf{3D Embedded Cosine Similarity:}\label{sec:3dsim_ae}
Alternatively, the embedded features of the unit cells in a given dataset can be extracted using deep learning models as simple as an autoencoder, a dimension reduction technique that compresses, \ie{}, encodes, complex shapes into vectors. Once such a model has been trained, an embedding-based shape similarity metric can be defined as the similarity between the vector representations of unit cells, much like the 2D descriptor-based distance earlier.

Here we also leverage point clouds, which are growing as a scalable and powerful representation for 3D deep learning~\cite{Guo2020deeppoints}. We utilize a point cloud autoencoder provided by Achlioptas~\etal{}~\cite{achlioptas18ae} with the Earth Mover's distance as the reconstruction loss. Our 3D dataset (described in Sec.~\ref{sec:3dgen}) is split into training, test and validation sets by 70\%, 15\%, and 15\%, respectively, and a grid search is performed to decide the hyperparameters: 64-dimensional embedded vectors for each unit cell, a learning rate of 0.0005 and batch size of 32. After training the model for 120 epochs, we can then take the cosine similarity between the embedded vector representations of any two unit cells as the shape metric. In our 3D experiment (Sec.~\ref{sec:3ddpp}), we compare the diverse subsets obtained using this embedded feature approach against those using the Hausdorff distance.

\subsection{Determinantal Point Processes for Joint Diversity in Two Spaces}\label{sec:dpp}
With a similarity kernel matrix $L$, we can now measure the diversity of a dataset using Determinantal Point Processes (DPPs), which are models of the likelihood of choosing a diverse set of items. They have been used for set selection in ML, \eg{}, diverse pose detection and information retrieval~\cite{kulesza2012determinantal,kulesza2011k}, and recently in ranking design ideas based on diversity and quality~\cite{Ahmed2017ranking}.
Viewed as joint distributions over the binary variables that indicate item selection, DPPs capture negative correlations. 
This means that, intuitively, the determinant of $L$ is related to the volume that the set covers in a continuous space. In other words, the larger the determinant, the more diverse the set.

To model our data, we construct DPPs through L-ensembles~\cite{borodin2009determinantal}, using a positive semi-definite matrix $L$ to define a DPP. Hence, given the full unit cells dataset of size $N$, which we denote as ground set $G$, DPPs allow us to find the probability of selecting any possible subset $M$ of unit cells as:
\begin{equation} \label{eq:dppk}
    \mathbb{P}(M) = \frac{det(L_M)}{det(L+I)},
\end{equation}
where $L_M \equiv [L_{ij}]_{ij \in M}$ is the submatrix of $L$ with entries indexed by elements of the subset $M$, and $I$ is a $N \times N$ identity matrix. The probability of a set containing two items increases as the similarity between them decreases. Therefore, the most diverse subset of any size has the maximum likelihood $\mathbb{P}(M)$, \ie{}, the largest determinant.
For a fixed subset size, the denominator can be ignored when maximizing the diversity via an algorithm such as the one described in Sec.~\ref{sec:ranking}.

Unlike submodular clustering approaches, DPPs only require the similarity kernel matrix $L$ as an input, and do not explicitly need the data to be clustered or a function that models diversity to be defined. This also makes them more flexible, since we only need to provide a valid similarity kernel, rather than an underlying Euclidean space or clusters.

For METASET, we calculate two different similarity values~---~one in shape space and another in property space~---~between any two unit cells. Hence, for all the unit cells combined, we have one kernel matrix corresponding to each of the two spaces. 
In order to measure the joint coverage in both spaces, we take a weighted sum of the two matrices, thus also allowing the trade-off between diversifying in shape or property space:
\begin{equation} \label{eq:jointkernel}
    L = (1-w) \cdot L_P + w \cdot L_S,
\end{equation}
where $L$, $L_P$ and $L_S$ are, respectively, the joint, property and shape similarity kernels, and $w$ is a weight parameter can be varied between 0 and 1. By adding the two kernels, we assume that the total similarity between two unit cells is the weighted average of how similar they are in the shape and property spaces.

While it is possible to combine two kernel matrices in many ways, we choose this formulation for two reasons. First, the weighted sum of two positive semi-definite matrices is also positive semi-definite, which is a pre-requisite for a DPP kernel. Second, it allows us to control the amount of diversity in both spaces, as well as to frame the later subset selection problem as multi-objective one, using a single tuning parameter $w$. We conducted multiple experiments on simulated data with easy-to-verify coverage metrics and found that this approach is effective in capturing diversity in both spaces. For brevity, we have not included these experiments here but directly report and discuss the results using joint kernels for metamaterials in Secs.~\ref{sec:2ddpp} and~\ref{sec:3ddpp}.

\begin{algorithm}
 \KwData{Ground set $G$ of size $N$ of all unit cells }
 \KwResult{Subset $M$ of size $N_M$}
 Calculate shape and property similarity kernels, $L_S$ and $L_P$\;
 Calculate joint similarity kernel $L$\;
 Find subset $M$\;
 \hspace{5mm}$M \gets \emptyset$\; %\label{alg:emptyS}
  \While{$|M| \neq N_M $}{
  Pick an item $G_i$ that maximizes $\delta f(M \cup {i})$\;
  $M = M \cup \{G_i\}$\;
  $G = G - G_i$\;
 }
 \Return{$M$}\; 
 Use $M$ as input to downstream task such as data-driven design or machine learning\;
 \caption{METASET algorithm. After calculating the similarity kernels, a polynomial-time greedy maximization of the gain on the weighted combination of diversity in shape and property spaces is performed. The output is a subset of unit cells such that the joint diversity is maximized.}
\label{alg:alg_metaset} 
\end{algorithm}

\subsection{Algorithm for Optimizing Diversity}\label{sec:ranking}
Optimizing the diversity of a subset $M$ in two spaces is an inherently multi-objective problem that can be accomplished by maximizing the log determinant of the joint similarity kernel, \ie{}, $f=\log [det(L_M)]$. Note that the log determinant of a positive semi-definite matrix is monotonically non-decreasing and submodular.
In general, finding the set of items that maximizes a submodular diversity function is NP-Hard. When solving such problems, a well-known limit due to Feige~\cite{feige2011maximizing} is that any polynomial-time algorithm can only approximate the solution up to $1-\frac{1}{e}\approx 67$\% of the optimal.

However, this is where choosing a submodular function $f$ as the objective comes in handy. It turns out that greedily maximizing this function is guaranteed to achieve the optimality bound~\cite{feige2011maximizing}. 
We use this property to substantially accelerate diversity optimization using a scalable greedy algorithm~\cite{nemhauser1978analysis}, which has theoretical approximation guarantees and is widely used in practice.
At each step, the algorithm picks an item, \ie, a unit cell, that provides the maximum marginal gain in the objective function (lines 5-8 in Algorithm \ref{alg:alg_metaset}). This makes greedy maximization of diversity the best possible polynomial-time approximation to an otherwise NP-Hard problem.

\section{METASET in Data-Driven 2D Metamaterials Design}\label{sec:2dcase}
Selecting a diverse and economical dataset prior to design can augment the performance and results of any data-driven algorithm. In this section, we demonstrate that this improvement can be achieved by adding METASET to existing data-driven frameworks with little extra cost (Fig.~\ref{fig:flow}) by designing 2D aperiodic mechanical metamaterials that meet desired displacement profiles and constraints on the connectivity of neighboring unit cells. 
Given a 2D dataset of unit cells from our previous work (briefly described in Sec.~\ref{sec:2dgen}), we use METASET to select several subsets with differing sizes and diversity scores (Sec.~\ref{sec:2ddpp}). 
By employing these subsets to assemble full structures, we study the effects of subset size and diversity on the search process and final designs (Sec.~\ref{sec:2ddesign}). To emphasize that our diverse selection methodology is an advantageous addendum to any data-driven method, we perform the designs with two existing approaches~---~genetic algorithm for an illustrative example, and a two-stage method for a more complex design motivated by practical applications (Sec.~\ref{sec:2ddesign_MRF}).

\begin{figure}[t]
    \centering
    \begin{subfigure}[b]{0.25\columnwidth}
        \centering
        \includegraphics[width=1\columnwidth]{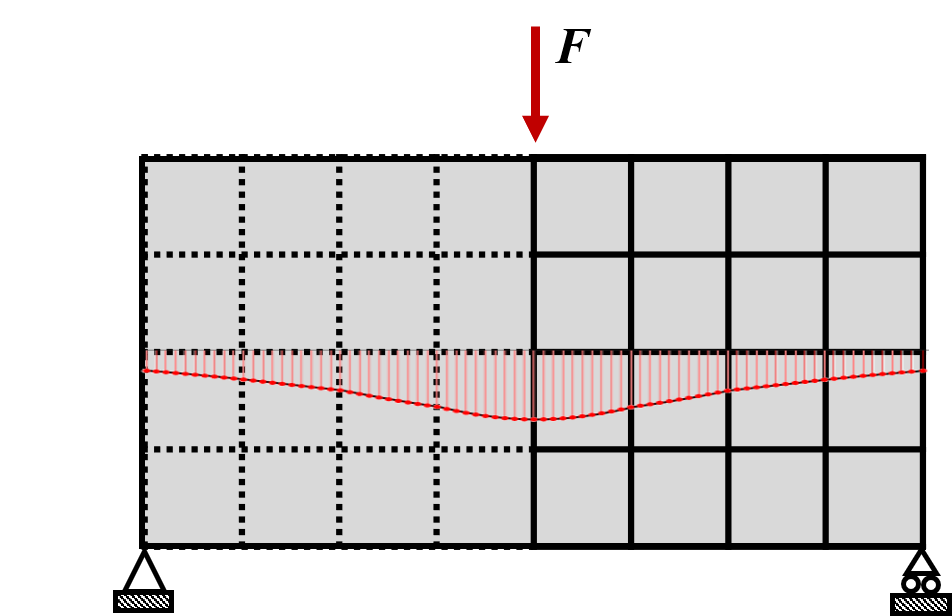}
        \caption{Classic MBB beam (Sec.~\ref{sec:2ddesign})}
        \label{fig:2dsetting_mbb}
    \end{subfigure}\qquad
    \begin{subfigure}[b]{0.6\columnwidth}
        \centering
        \includegraphics[width=1\columnwidth]{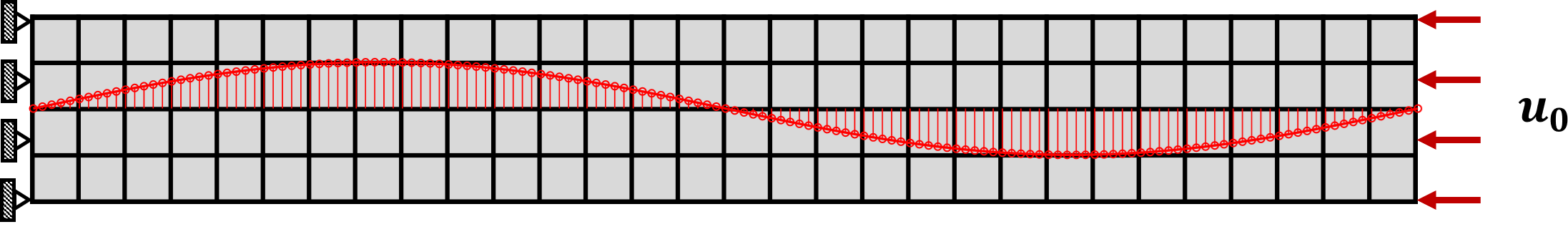}
        \caption{Cantilever (Sec.~\ref{sec:2ddesign_MRF})}
        \label{fig:2dsetting_sine}
    \end{subfigure}
    \caption{Problem settings of the 2D examples, both of which should achieve the target displacement profiles shown in red.}
    \label{fig:2dsettings}
\end{figure}

The design settings, a classic MBB beam and a cantilever, along with the boundary conditions and target displacement profiles (red curves) are shown in Fig.~\ref{fig:2dsettings}. The design objective for both is to minimize the mean squared error (MSE) between the target and achieved displacement profiles. These types of structures, which require spatially varying elastic behavior and therefore benefit from aperiodic configurations and data-driven methods, have been a growing focus in recent research, with applications such as soft robotic grippers and biomedical devices~\cite{Schumacher2015,Mirzaali2018softdevice}. 
We deliberately choose these since spatially varying properties are difficult to obtain using conventional methods, particularly when the objective is dependent on the relative spatial distribution of properties rather than an absolute performance value like compliance.

To support this claim, we attempted to benchmark the performance of a conventional TO approach based on the Solid Isotropic Material with Penalization (SIMP) scheme~\cite{Bendse2004simp} for the MBB problem (Eq.~\ref{eq:ga_problem}), whose sensitivities can be derived using adjoint analysis. Each unit cell is discretized into $50\times 50$ quadrilateral finite elements, and the density of each element is treated as a design variable, $\rho_e \in [0,1]$, where the goal is to converge as close to 0 (void) or 1 (solid) as possible. To eliminate mesh dependency, a sensitivity filter with a radius of 2 is applied. For combinations of different penalty factors, $p\in \{1,3\}$, and volume fraction constraints, $V\in \{0.50, 0.75, 1.0\}$, we minimize the MSE using the Method of Moving Asymptotes (MMA)~\cite{Svanberg1987mma} and the same stopping criteria. All results are infeasible, however, with high MSE ranging from 4.69 to 6137.08 and numerous intermediate densities (more than 95\% of the elements). This underscores the need for more advanced approaches like data-driven design, which have successfully achieved target spatially varying behavior~\cite{Schumacher2015,Coulais2016combinatorial,Zhu2017twoscale,Wang2020smo}. We will leverage two such approaches in the following sections, since the goal of this paper is not to propose new design methods but to select diverse subsets which provide salient advantages to any existing data-driven design framework.

\subsection{Generation of 2D Unit Cells via Topology Optimization and Perturbation}\label{sec:2dgen}
In~\cite{Bostanabad2019gagp,Wang2020smo}, we previously proposed using a combination of TO and stochastic shape perturbation to generate a large dataset of 2D unit cells. To initialize the dataset, we ran density-based TO for each uniformly sampled target property, the components of homogenized elastic tensors, and then iteratively perturbed the shape of the unit cells with the most extreme or uncommon properties. By doing so, we created a dataset of 88,000 unit cells that covered a relatively large property space within reasonable computational cost. Note that we did not build this dataset with geometry in mind, leading to many similar shapes. Also, even though we aimed to fill the less populated regions of the property space by perturbing unit cells in those locations, there is a higher concentration of final unit cells with lower property values (the lower left corners in Fig.~\ref{fig:2d_project_prop}), indicating that the dataset is somewhat imbalanced. For details, please see~\cite{Wang2020smo}.

Before applying METASET, we preprocess the data by randomly sampling unit cells from the original dataset that have a volume fraction greater than 0.70, resulting in 17,380 unit cells. This fraction was chosen so that the chosen unit cells are less likely to have very thin features, which makes them more feasible for manufacturing. 
Additionally, when computing shape diversity, if unit cells occupy very different volume fractions, a diverse subset is more likely to be dominated by flimsy, low density structures, whose shapes have the least probability of overlap with other unit cells.
However, as we will show with the design examples, this preprocessing does not impede the chances of designing well-connected structures that met the targets quite well.

\subsection{Diverse 2D Unit Cells}\label{sec:2ddpp}
For the dataset of 17,380 2D unit cells, which we now refer to as the full or ground set $G$, we calculate the property and shape similarity matrices, $L_P$ and $L_S$, respectively, as described in Sec.~\ref{sec:similarity}. Taking their weighted sum forms the joint DPP kernel matrix $L$ (Sec.~\ref{sec:dpp}), whose determinant, $det(L_M)$, scores the diversity in both spaces. To explore this, we rank several subsets using the greedy algorithm from Sec.~\ref{sec:ranking} by varying their sizes, $N_M$, and kernel weights, $w$. From the results, we can make three observations:
\begin{enumerate}
    \item By increasing $w$, we shift from ranking a subset based on diversity in the property space alone, to a mixture of both spaces, and to the shape space only. In essence, the trade-off between shape and property diversity can be easily controlled.
    \item The correlation coefficient between the shape and property diversity scores of 1,000 random subsets of size five is 0.0047. Similar near-zero correlation is found for other set sizes too. In addition, the correlation between the shape and property similarity values of 100,000 random pairs of unit cells is $-0.0024$. Therefore, our assumption that the joint similarity can be modeled as a weighted sum is appropriate.
    \item By observing the joint diversity score of the subsets as more items, \ie{} unit cells, are added, we find that the gains in shape and property diversities saturate at approximately $N_M=20$. Thus, a very small number of unit cells are sufficient to cover both spaces.
\end{enumerate}

\begin{figure}[ht]
    \centering
    \begin{subfigure}{0.8\columnwidth}
        \includegraphics[width=1\columnwidth]{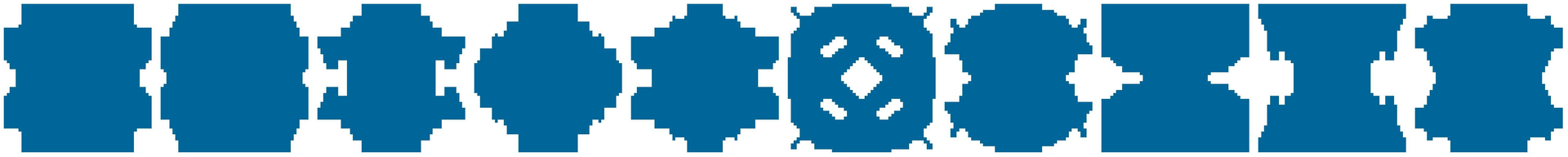}
        \caption{Subset diverse in property space ($w=0$)}
        \label{fig:2d_0}
    \end{subfigure}
    
    \begin{subfigure}{0.8\columnwidth}
        \includegraphics[width=1\columnwidth]{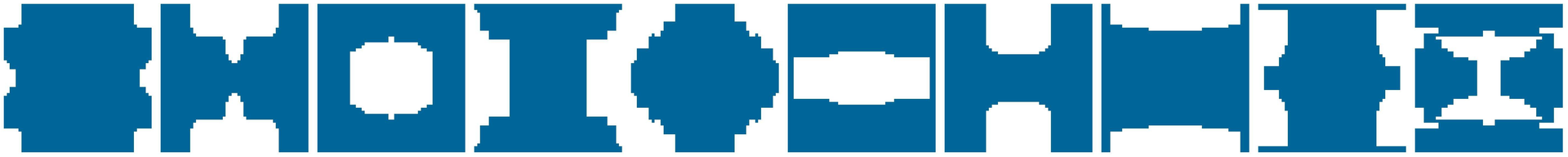}
        \caption{Subset diverse in shape and property spaces ($w=0.5$)}
        \label{fig:2d_05}
    \end{subfigure}
    
    \begin{subfigure}{0.8\columnwidth}
        \includegraphics[width=1\columnwidth]{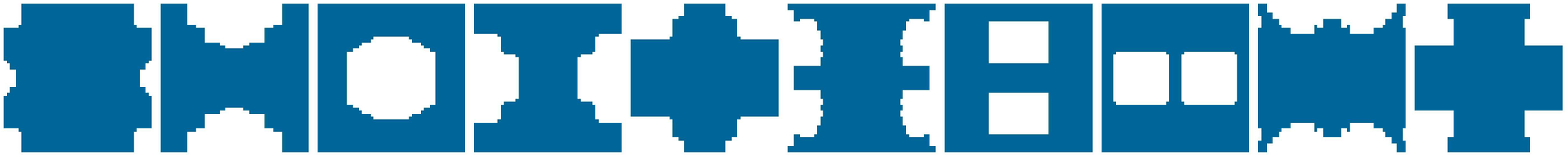}
        \caption{Subset diverse in shape space ($w=1$)}
        \label{fig:2d_10}
    \end{subfigure}
    
    \caption{Examples of 2D unit cells from the diverse subsets used in the cantilever and MBB design problems.}
    \label{fig:2d_sets}
\end{figure}

Ten example unit cells from the subsets with $w\in\{0,0.5,1\}$ are shown in Fig.~\ref{fig:2d_sets}, where the subset optimized for only shape diversity (Fig.~\ref{fig:2d_10}) displays the most variety of topologies compared to the subset diverse in only properties (Fig.~\ref{fig:2d_0}). Meanwhile, the balanced subset contains a mixture of unit cells akin to both extreme sets (Fig.~\ref{fig:2d_05}). This may be counter-intuitive since similar shapes should have similar mechanical properties. However, note that upon close inspection, the property diverse unit cells exhibit tiny features that lead to low effective elastic property values. Such small details in the shape may lead to a larger change according to the physical simulations and the property similarity metric, \ie{}, the Euclidean norm. 

\begin{figure}[hb]
    \centering
    \begin{subfigure}{0.3\columnwidth}
        \includegraphics[width=1\columnwidth]{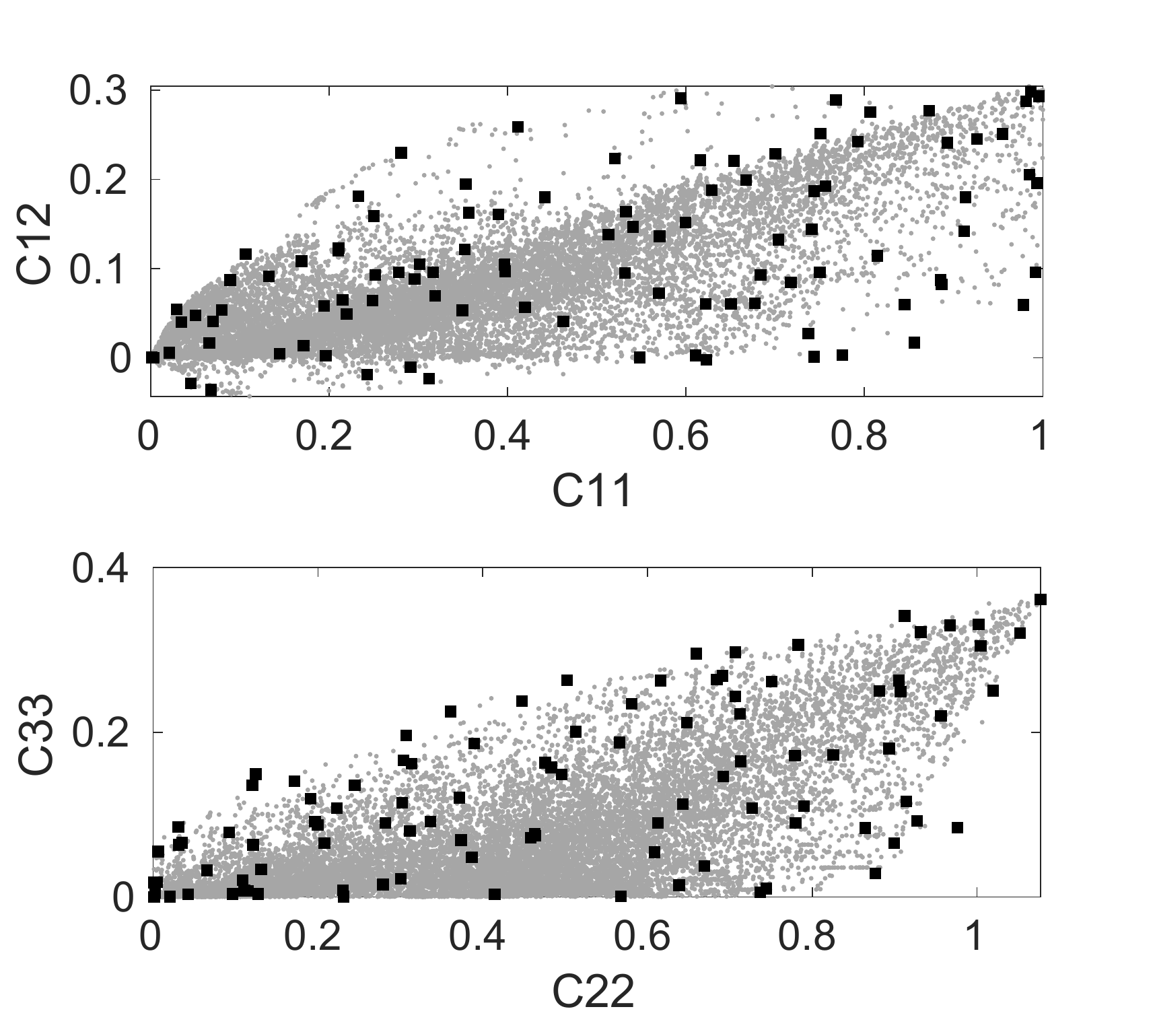}
        \caption{Property diverse samples}
        \label{fig:2d_project_prop_a}
    \end{subfigure}
    \begin{subfigure}{0.3\columnwidth}
        \includegraphics[width=1\columnwidth]{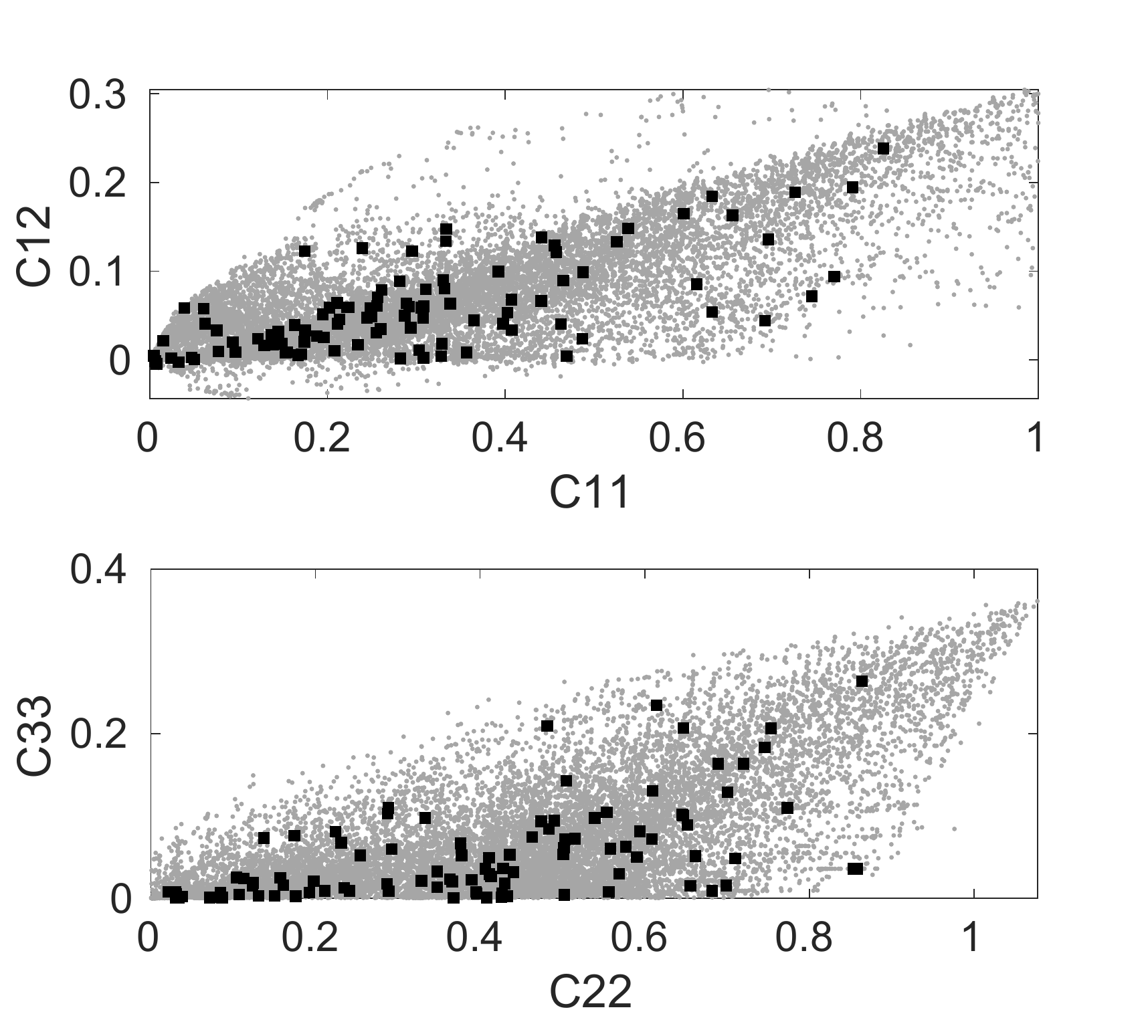}
        \caption{Random samples}
        \label{fig:2d_project_prop_b}
    \end{subfigure}
    \begin{subfigure}{0.3\columnwidth}
        \includegraphics[width=1\columnwidth]{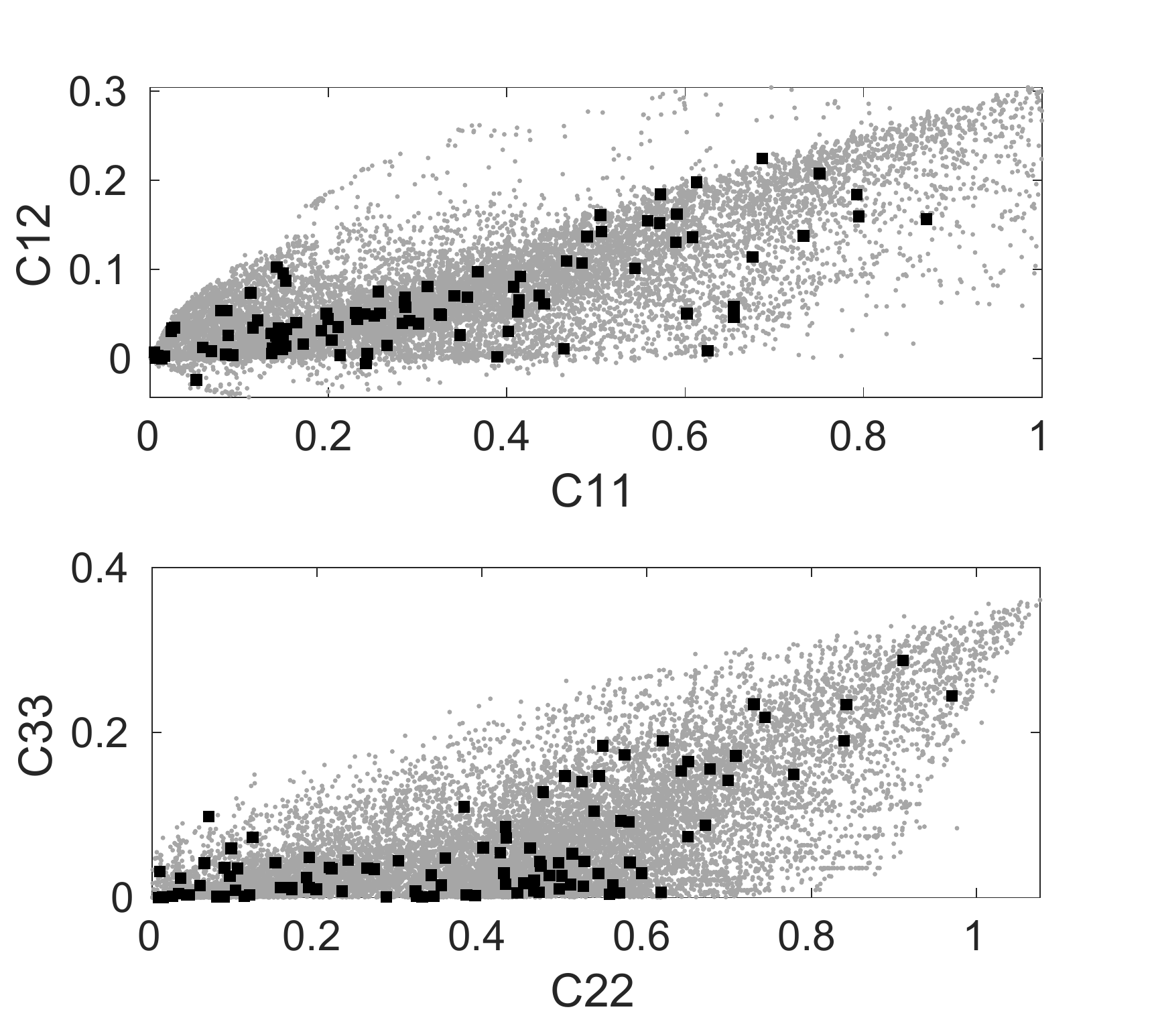}
        \caption{Shape diverse samples}
        \label{fig:2d_project_prop_c}
    \end{subfigure}
    
    \caption{The property space of the 2D unit cell subsets optimized for property and shape diversity, and a randomly sampled set, plotted against the full dataset. We observe that property diverse subsets cover the space well, hence it is more likely to have unit cells near any target property combination.}
    \label{fig:2d_project_prop}
\end{figure}

Comparing the properties of the unit cells in diverse subsets to the ground and randomly sampled sets (Fig.~\ref{fig:2d_project_prop}), we can confirm that the property diverse subsets cover all regions of the original property space, even the sparsely populated areas. As expected, the shape diverse subset does not do as well, and the random subset contains tight clusters in certain areas. Along with the observation that the diversity scores as well as the similarity values in the shape and property spaces are essentially uncorrelated, these findings confirm that the formulation of the joint kernel $L_M$ as a weighted linear sum (Eq.~\ref{eq:jointkernel}) is effective for controlling the amount of diversity in either space.

Finally, the result that only 20 unit cells is needed to cover the shape and property spaces is quite interesting since a main tenet of data-driven design thus far is that ''more is better''~---~larger datasets provide more candidates from which we can choose compatible unit cells. So, to explore the impact of the subset size on the data-driven approach, we selected the top 20 as well as top 100 ranking unit cells from each subset to move on to the next step: full structure assembly.

\subsection{Illustrative Study on the Effects of Size and Diversity}\label{sec:2ddesign}
We begin by designing a relatively simple classical example from the TO field, the MBB beam, such that its horizontal centerline conforms to the red curve when loaded with a vertical force $F$ (Fig.~\ref{fig:2dsetting_mbb}). Due to the structural symmetry, we only need to design the right half of the beam with $4\times 4$ unit cells, outlined by the solid black lines. The full structure can then be obtained by reflecting over the vertical centerline.
Using subsets of unit cells with varying sizes and levels of diversity for metamaterials design using global optimization, we can elucidate 1) the effect of subset size on the search algorithm's efficiency, and 2) the impact of diversity on the final design performance as well as the compatibility of neighboring unit cells. 
We choose the following diverse subsets using METASET:
\begin{itemize}
\setlength{\itemsep}{0pt}
    \item $P_{20}$: Property diverse subset of size 20
    \item $SP_{20}$: Shape and property diverse subset of size 20
    \item $S_{20}$: Shape diverse subset of size 20 diverse
    \item $P_{100}$: Property diverse subset of size 100
    \item $SP_{100}$: Shape and property diverse subset of size 100
    \item $S_{100}$: Shape diverse subset of size 100.
\end{itemize}
In addition, we utilize these sets, which are not diverse and are not selected by our method, as baselines:
\begin{itemize}
\setlength{\itemsep}{0pt}
    \item $R_{20}$: Random subset of size 20
    \item $R_{100}$: Random subset of size 100
    \item $G$: Full dataset of size 17,380.
\end{itemize}

To design the MBB beam, we pass each of the datasets to a global optimization method, which for this example is a single objective genetic algorithm. Although the approach is simple, we chose it to focus on illustrating the effects of subset size and diversity on the final results. It also allows us to restrict our design to the discrete choice of unit cells in our subsets, whereas most gradient-based algorithms for data-driven metamaterials design map continuous design variables to the nearest existing, or interpolated, unit cell in dense databases \cite{Schumacher2015,Zhu2017twoscale}.

Specifically, the genetic algorithm is used to select the combination of unit cells from each given dataset that minimizes the MSE between the achieved and target displacement profiles. In addition, since detached neighbours are not desirable, we add a compatibility constraint by requiring that the number of disconnected unit cells, $N_{dc}$, in the full structure be equal to zero. The optimization problem is formulated as:
\begin{equation}\label{eq:ga_problem}
\begin{aligned}
    & \underset{\bm{l}}{\text{minimize}}
    & & \frac{1}{n} \lVert \bm{u}(\bm{l})-\bm{u}_t \rVert _2^2 \\
    & \text{subject to}
    & & \bm{K}(\bm{l})\bm{U} = \bm{F},\\
    & & & N_{dc}(\bm{l})=0,\\
    & & & l_i \in \{1,2,\cdots,N_M\}, \quad i=1,2,\ldots,N_f,
\end{aligned}
\end{equation}
where $\bm{u}$ is the displacement of $n$ nodes located on the centerline of the structure, $\bm{u}_t$ is the discretized target displacements, $\bm{K}$ is the global stiffness matrix, and $\bm{U}$ and $\bm{F}$ are global displacement and loading vectors, respectively. The number of unit cells in the given dataset is $N_M$ while the number in the full structure is $N_f$, and $\bm{l}=[l_1,l_2,\ldots,l_{N_f}]^T$ is a vector of the indices of the chosen unit cells.

\begin{figure}[ht]
    \centering
    \includegraphics[width=0.55\columnwidth]{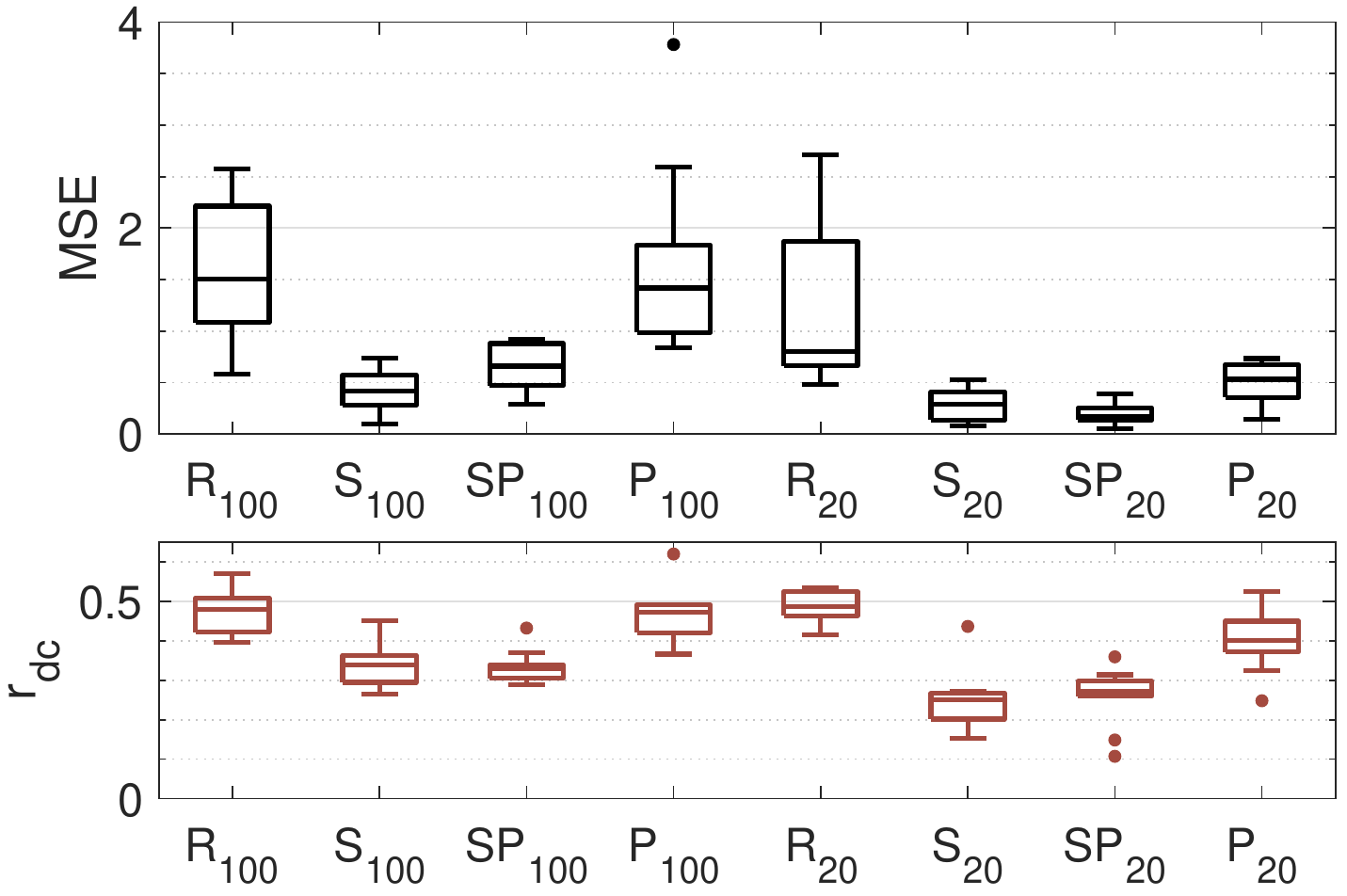}
    \caption{The final objective values (MSE) and ratios of disconnectivity ($r_{dc}$) of 10 runs per subset. Lower values are better. The best overall MSE is obtained by $SP_{20}$ and $S_{20}$, and the best $r_{dc}$ by $S_{20}$ and $SP_{20}$.}
    \label{fig:2d_case_bar}
\end{figure}

Due to the stochasticity of genetic algorithms, we run the optimization ten times for each dataset and report the MSE of the final topologies in Fig.~\ref{fig:2d_case_bar}. In addition, we show a measure of the connectivity of the final structure: the mean ratio of disconnected pixels on the boundaries of touching unit cells, $r_{dc}$. Similar to $N_{dc}$ in the constraint (Eq.~\ref{eq:ga_problem}), a fully compatible structure should have $r_{dc}$ as zero. The averages of these results are also disclosed in Table~\ref{table:2d_mean}.

\begin{table}[ht]
    \caption{Means of the final results for the MBB example, with the lowest values in \textbf{bold}.}
    \centering
    \begin{tabular}{@{}llllllllll@{}}
    \toprule
                    & $G$      & $R_{100}$ & $S_{100}$ & $SP_{100}$ & $P_{100}$ & $R_{20}$ & $S_{20}$ & $SP_{20}$ & $P_{20}$ \\ \midrule
    MSE         & 1.3E+18 & 1.5341    & 0.4278    & 0.6454     & 1.6648    & 1.2395   & 0.2865   & \textbf{0.2017}    & 0.4926   \\
    $r_{dc}$    & 0.5184   & 0.4770    & 0.3406    & 0.3347     & 0.4653    & 0.4836   & \textbf{0.2488}   & 0.2578    & 0.3996   \\ \bottomrule
    \end{tabular}
\label{table:2d_mean}
\end{table}

When given the baseline full dataset, $G$, the genetic algorithm is overwhelmed and not able to find any designs with satisfactory MSE (see the high values in Table~\ref{table:2d_mean}), even failing to meet the compatibility constraint in one run. This can be attributed to a vast search space since the number of possible unit cell combinations grows exponentially as the size of the dataset increases. A larger set may also contain more redundant shapes or properties that contribute little to diversity, exacerbating the search challenge and possibility of local optima.
Conversely, every run using the 20- and 100-item subsets satisfy the design requirements (Fig.~\ref{fig:2d_case_bar} and Table~\ref{table:2d_mean}). These include the baseline random subsets selected without our method, which obtain reasonable performance and connectivity due to the reduced search space. The values of MSE and $r_{dc}$ using random subsets, however, vary widely.
In fact, our results highlight that \textit{smaller yet diverse} subsets more consistently outperform all other sets under the same search algorithm and termination criteria. Notably, the lowest mean MSE is reached by the small $SP_{20}$ and $S_{20}$ sets. 
Moreover, the best connected structures, \ie{}, those with lowest $r_{dc}$, result from the diverse subsets that consider shape, \ie{}, $S_{20}$ and $SP_{20}$. 
We remark that our optimization problem only constrains the number of disconnected unit cells and does not explicitly minimize $r_{dc}$. Therefore, the shape diverse results naturally attain higher connectivity.

\begin{figure}[ht]
    \centering
     \begin{subfigure}{0.7\columnwidth}
     \centering
        \includegraphics[width=1\columnwidth]{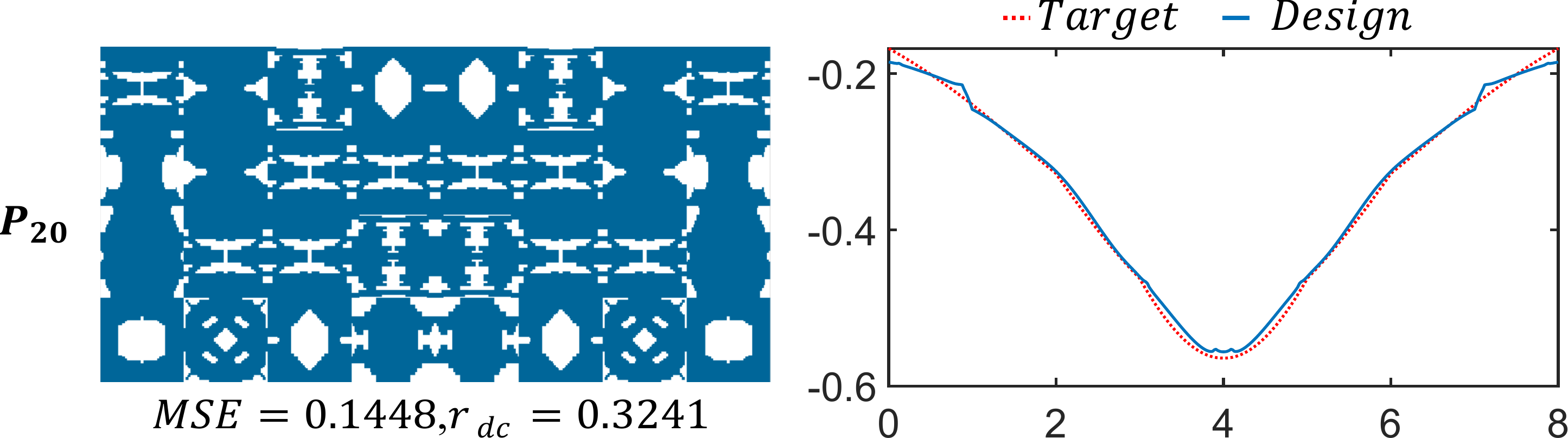}
        \caption{Using property diverse subset of size 20 ($P_{20}$), the unit cells are connected, but by small features.}
        \label{fig:cons_p20}
    \end{subfigure}
    
    \begin{subfigure}{0.7\columnwidth}
    \centering
        \includegraphics[width=1\columnwidth]{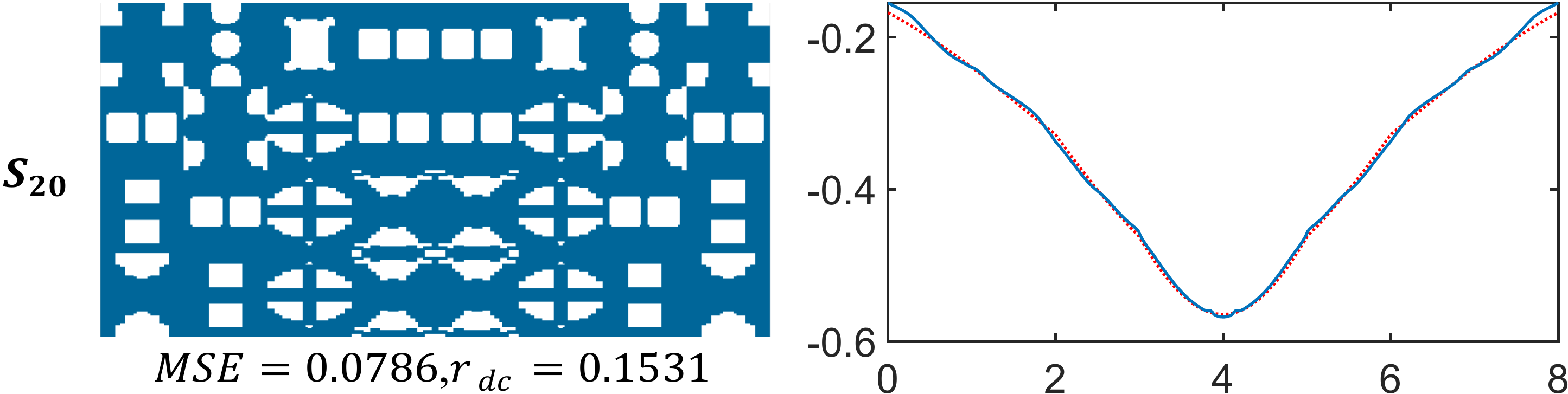}
        \caption{Using shape diverse subset of size 20 ($S_{20}$), we observe superior connectivity between neighboring unit cells.}
        \label{fig:cons_s20}
    \end{subfigure}

    \caption{Final topologies and displacement profiles of the classic MBB beam example with the lowest MSE out of 10 runs using 20-item diverse sets. The full structure after symmetry is shown.}
    \label{fig:2dphi}
\end{figure}

 Fig.~\ref{fig:2dphi} shows the final topologies and optimal displacement profiles of the runs that achieve the minimum MSE for two datasets. As expected from the worse performance and compatibility, the designs using the full dataset $G$ (not pictured) contain disconnected and oddly matched unit cells. 
In a similar vein, the high $r_{dc}$ for property diverse sets correspond to mediocre connectivity, as shown by the $P_{20}$ result in Fig.~\ref{fig:cons_p20}, where neighbors are linked by tiny features. 
This can be associated with the observation in Sec.~\ref{sec:2ddpp} that METASET tends to include unit cells with small features as it maximizes property diversity, leading to subsets with less compatible unit cells. 
With shape diverse subsets, however, the final designs possess excellent compatibility, such as in Fig.~\ref{fig:cons_s20}, further enforcing the advantages of shape diversity.

Although our constrained genetic algorithm provides satisfactory designs, we must point out that our goal is not to introduce new design methods; this global method was implemented to showcase the impact of subset size and diversity. While more elegant optimization techniques would be better suited for practical applications, we nevertheless believe that the insights gained from this study~---~that selecting diverse subsets can accelerate and benefit metamaterial design~---~can be generalized to other data-driven methods, such as the one in the next section.

\subsection{Additional Study with a Complex Metamaterial Structure}\label{sec:2ddesign_MRF}
In the previous section, a simple example using genetic algorithm demonstrated that data-driven metamaterials design can benefit from small and diverse subsets of unit cells. To validate that this is also true for more sophisticated algorithms and designs, we now test the same hypothesis by combining our diverse subsets with an advanced optimization method we proposed in~\cite{Wang2020smo}, which is described briefly below. 
Here we design a cantilever composed of $4 \times 30$ unit cells to achieve a sine-wave shape when a prescribed displacement boundary condition is imposed (Fig.~\ref{fig:2dsetting_sine}). As opposed to the MBB beam, the spatially varying behavior of the cantilever is designed to deform in opposite directions in the left and right halves, and we expect that different regions in the structure will require distinctly contrasting properties. 
The prescribed boundary instead of a point load poses an additional challenge. The closest problem to this that has been addressed by traditional TO methods is the compliant mechanism design, which aims to control the ratios between output and input displacements or forces by minimizing the displacement at a particular node. To obtain feasible mechanism designs, however, Deepak~\etal{} found in~\cite{deepak2009comparative} that it is necessary to assume a force-displacement relationship, \ie{}, a spring, at that output node. In contrast, our problem minimizes the MSE over all nodes along the centerline. Since adding a spring at each of those would significantly deviate from our problem setting, conventional design methods are not plausible.

Due to the difficulty of this problem, or indeed any realistic metamaterials design, searching over larger datasets to locate compatible unit cells while meeting the desired performance is also expensive or even intractable. In our case, we are only able to use the smaller diverse subsets $S_{20}$, $SP_{20}$ and $P_{20}$ introduced earlier, as well as baseline random subsets $R_{20}$. 
Since there are 120 macro-elements in the cantilever, this still means that there are $120^{20}$ possible combinations of unit cells for each subset.

For this example, we follow our two-stage optimization framework~\cite{Wang2020smo}, wherein inverse TO is utilized in the first stage to determine the macroscale property distribution, and combinatorial optimization based on weighted graphs is used in the second stage to assemble unit cells that meet the target properties with compatible boundaries.
Specifically, we define the following optimization problem for the first stage:
\begin{equation}\label{eq:Inverse_problem}
\begin{aligned}
    & \underset{\bm{C_{e}}}{\text{minimize}}
    & & \frac{1}{n} \lVert \bm{u}(\bm{l})-\bm{u}_t \rVert _2^2 \\
    & \text{subject to}
    & & \bm{K}(\bm{C_{e}})\bm{U} = \bm{F},\\
    & & & -\phi(\bm{C_{e}}) \leq 0.\\
\end{aligned}
\end{equation}
Compared to the problem solved via genetic algorithm in the previous section (Eq.~\ref{eq:ga_problem}), this inverse property design directly uses the element stiffness matrix $\bm{C_{e}}$ as design variables, which are constrained by the signed L2 distance field $\phi$ of the property space of the full subset $G$. This inverse problem can be efficiently solved with MMA~\cite{Svanberg1987mma}.

After obtaining the optimized macro-property distribution, we construct a grid-like weighted graph with each node representing an element in the macrostructure, and with edges connecting neighbouring unit cells. We can then view the assembly problem as selecting an index from the given subset to label each node in the graph. The Euclidean distance to the target property is assigned as the nodal weight during this process, and the ratio of disconnectivity, $r_{dc}$ defined in the last section, is assigned as the edge weight for each pair of neighboring nodes. With this graph, we can use a dual decomposition Markov random field (DD-MRF) method~\cite{Komodakis2010} to efficiently find the optimal labels of the graph with the lowest sum of nodal and edge weights, thereby designing a full structure that meets the target properties and is well-connected.

\begin{figure}[ht]
    \centering
    \includegraphics[width=0.85\columnwidth]{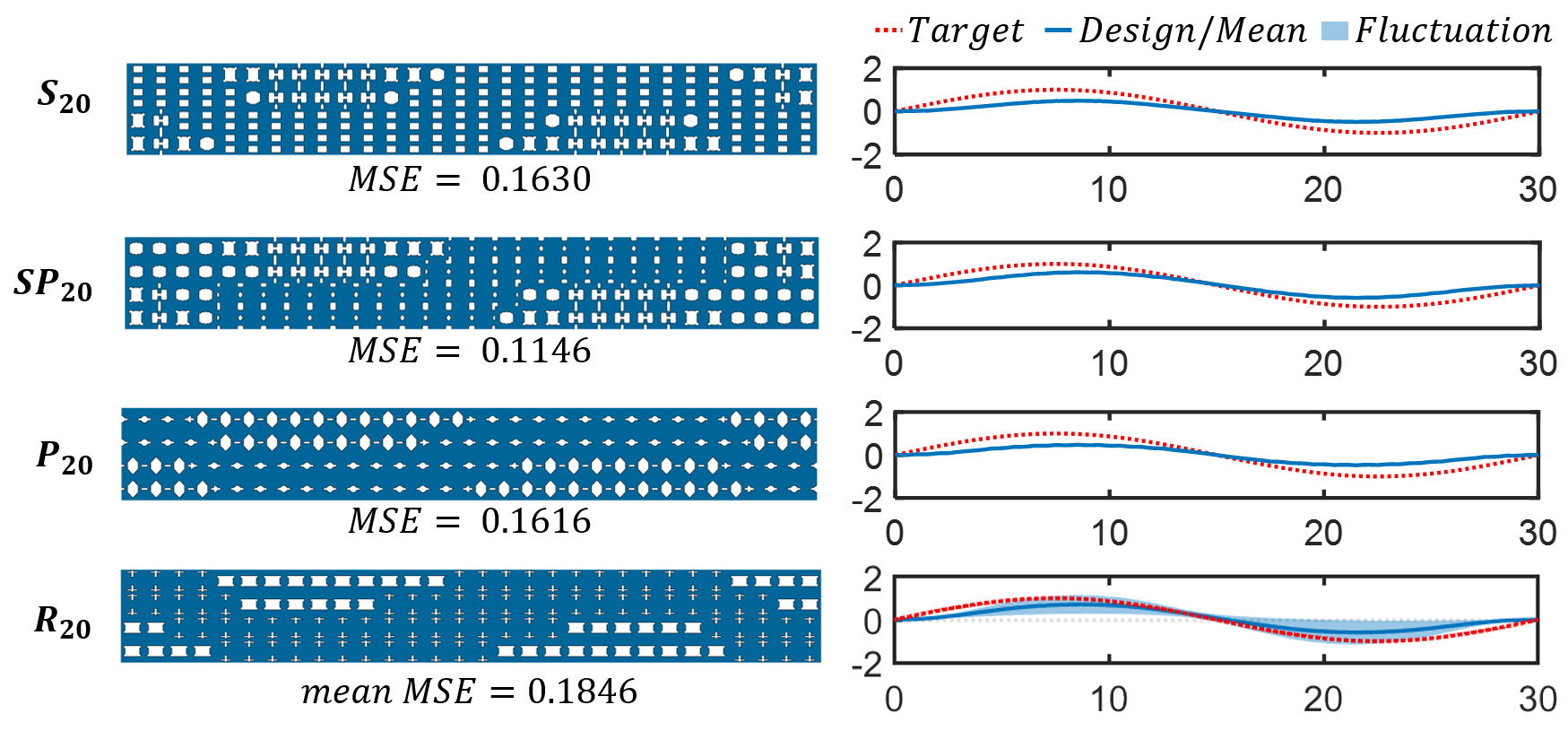}
    \caption{Optimized structures using different subsets, and their associated displacement profiles, for the cantilever example.}
    \label{fig:SineResult}
\end{figure}

Since the labeling problem for the graph is a complex combinatorial optimization process where a large candidate set of unit cells equates to an immense search space, a small subset is required for a higher efficiency. As aforementioned, we use three diverse subsets, $S_{20}$, $SP_{20}$, $P_{20}$, and five subsets randomly selected without METASET, $R_{20}$, each with 20 unit cells as the candidate sets for the second stage. The resulting full structures and their respective MSE values and displacement profiles are shown in Fig.~\ref{fig:SineResult}. We repeat the design using random subsets five times, then plot the mean displacement profile and depict the fluctuation of the results with the shaded area.

By virtue of our weighted graph method, all optimized designs have compatible boundaries. However, the subsets which account for shape diversity, \ie{}, $S_{20}$ and $SP_{20}$, include a wider variety of unit cells in the full structure. This can be credited to an observation we made in the previous MBB beam example, that a shape diverse set can provide more compatible pairs, rendering a larger feasible design space for the assembly problem. In addition, we note that although some random subsets can achieve relatively low MSE, this performance is not guaranteed; the mean MSE is still the worst overall. In constrast, the shape and property diverse subset $SP_{20}$ has the lowest MSE value. The reason is that, even with small subsets, shape diversity provides better compatibility while property diversity helps to achieve the target property distribution. This is again in line with our findings that a small yet diverse subset considering shape and properties is a boon for data-driven metamaterials design, and has exciting implications for future works.

\section{METASET for Discovery of Diverse 3D Unit Cell Families}\label{sec:3dcase}
Beyond selecting diverse subsets for direct use in design, another advantage of METASET is eliminating inherent bias by optimizing the diversity of a dataset. We demonstrate this with a 3D study, first introducing a new method based on periodic functions to generate families of unit cells with the same underlying structure but varying densities, which although fast creates a great number of overlapping shapes. Our goal in applying METASET to this 3D data is to sift through the overlaps to discover diverse sets of unique isosurface families, which can subsequently be leveraged for data-driven design or ML of, \eg{}, property prediction or generative models (Fig.~\ref{fig:flow}).

Triply periodic isosurface unit cells, whose symmetries follow those of crystal structures~\cite{Wohlgemuth2001bicontinuous}, are often used in 3D mechanical metamaterials design due to excellent surface area-to-performance ratios and manufacturability~\cite{Maskery2018tpms}. In addition, their representation as level-set functions allows the density of the unit cells to be easily manipulated for functionally-graded structures~\cite{Maskery2018tpms,Li2019tpms} and tailorable acoustic bandgaps~\cite{Abueidda2018bandgap}. A level-set function $f(x,y,z) = t$ is an implicit representation of geometry where the $t$-isocontour, \ie{}, the points where $f=t$, describes the surface of the structure, while the locations where $f<t$ are solid material, and void where $f>t$. Thus, by varying the isovalue $t$, an entire family of isosurface unit cells with graded densities can be extracted from one level-set function.

The most prevalent type of isosurfaces used in metamaterials design is a special subset known as Triply Periodic Minimal Surfaces (TPMS). However, only a few TPMS families have been used since their functions are complex to derive~\cite{Wohlgemuth2001bicontinuous}. For example, Maskery~\etal{} use six families in their design work~\cite{Maskery2018tpms}, while Li~\etal{} use four~\cite{Li2019tpms}. 
Moreover, it has not been investigated whether these few families cover the gamut of shapes and properties needed for design applications.
Suppose a researcher wishes to design a new functionally-graded 3D metamaterial by tuning the densities of isosurface functions, but does not know beforehand which families would best suit their application. Due to the computational expense of design in 3D, they may desire to select a smaller set of families that can then be used in their optimization method. In this section, we present METASET as a procedure to choose those families such that the resultant subset has large coverage over different properties and shapes. In doing so, we also demonstrate that METASET removes bias in datasets by maximizing diversity.

\subsection{Generation 3D Unit Cell Families using Level-Set Functions}\label{sec:3dgen}
Before selecting diverse families, we must first generate an initial pool to choose from. Thus, to build a large 3D dataset, we propose a new method to create isosurface families based on the level-set functions of crystallographic structure factors, which describe how particles are arranged in a crystal unit cell~\cite{ITC2010crystaltables}. In contrast to most unit cell generation methods, our approach here does not set targets in the property space or use TO, and different from TPMS functions, a larger variety of shapes can be found without complex derivations.

In crystallography, structures that are invariant under the same symmetry operations belong to the same space group, of which there are 230 for 3D structures. For the purposes of our work, we will focus on the 36 cubic groups, No. 195 through 230, to obtain our level-set functions. Experimentally, the space group of a crystal can be determined through, \eg{}, X-ray techniques, by scattering radiation off a lattice plane denoted by $(hkl)$, and then observing the diffraction pattern. These symmetric patterns have been analytically modeled as \textit{structure factors}, which are periodic functions of the form:
\begin{equation}\label{eq:SF_analytical}
    f_{group,(hkl)}(X,Y,Z) = A + i B,
\end{equation}
where $A = \cos{\big(hX+kY+lZ\big)}$, $B = \sin{\big(hX+kY+lZ\big)}$, $X = 2\pi x$, $Y = 2\pi y$, and $Z = 2\pi z$. The equations of these structure factors are listed in~\cite{ITC2010crystaltables} for all space groups and their allowable $(hkl)$.

\begin{figure}[htb]
    \centering
    \begin{subfigure}{0.4\columnwidth}
        \centering
        \includegraphics[width=1\columnwidth]{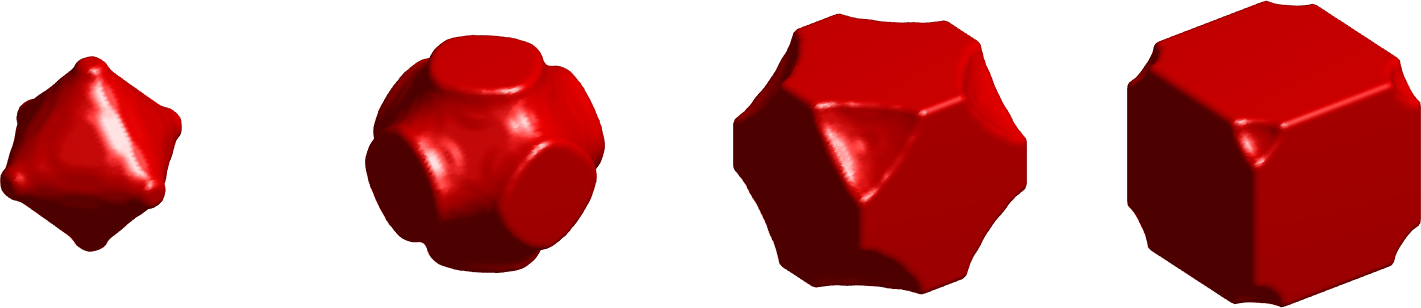}
        \caption{Family $A_{229,(001)}(X,Y,Z) \leq t$}
        \label{fig:iso_demo_tm}
    \end{subfigure}
    
    \begin{subfigure}{0.4\columnwidth}
        \centering
        \includegraphics[width=1\columnwidth]{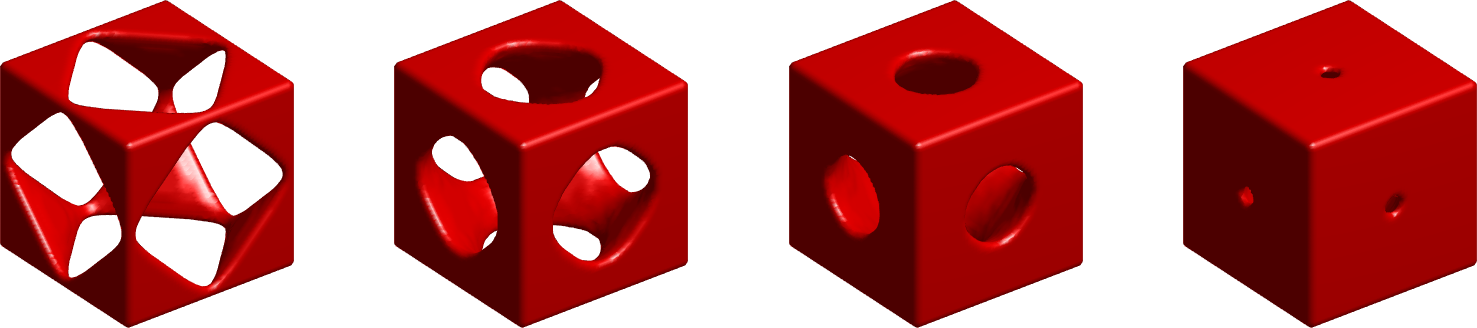}
        \caption{Family $A_{229,(001)}(X,Y,Z) \geq t$}
        \label{fig:iso_demo_tp}
    \end{subfigure}
    
    \begin{subfigure}{0.4\columnwidth}
        \centering
        \includegraphics[width=1\columnwidth]{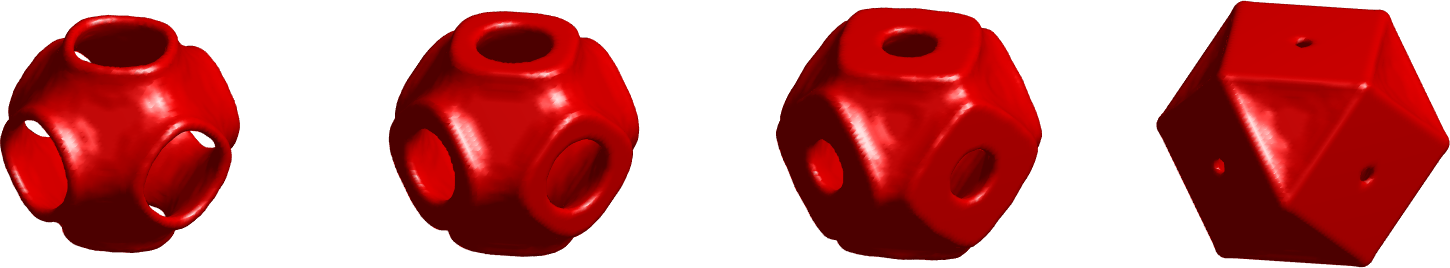}
        \caption{Family $A^2_{229,(001)}(X,Y,Z) \leq t^2$}
        \label{fig:iso_demo_tt}
    \end{subfigure}
    \caption{Examples of unit cells from isosurface families generated by the structure factor for space group No. 229 and $(hkl)=(001)$. The effect of increasing $t$ to create a family is shown from left to right.}
    \label{fig:iso_demo}
\end{figure}

We can split each structure factor into six isosurface families by separating $A$ and $B$ in Eq.~\ref{eq:SF_analytical} (inspired by~\cite{Wohlgemuth2001bicontinuous}), and converting them into level-set functions as follows:
\begin{equation}\label{eq:SF_forms}
\begin{aligned}
    A_{group,(hkl)}(X,Y,Z) &\leq t, \\
    A_{group,(hkl)}(X,Y,Z) &\geq t, \\
    A^2_{group,(hkl)}(X,Y,Z) &\leq t^2,
\end{aligned}
\end{equation}
and similarly for $B_{group,(hkl)}$. These, respectively, correspond to setting as solid material the function values that are less than $t$ (Fig.~\ref{fig:iso_demo_tm}), greater than $t$  (Fig.~\ref{fig:iso_demo_tp}), and in between $-t$ and $t$ (leading to a ''thin-walled'' structure; Fig.~\ref{fig:iso_demo_tt}).

Thus, instead of using the limited TPMS functions, we can use the structure factors of all 36 cubic space groups and their corresponding $(hkl)$ to generate a greater number of isosurface families for data-driven design. To ensure manufacturability, we also identify the feasible density range of each family by prohibiting internal voids and disconnected features, and eliminate families whose feasible range is $\rho_{max} - \rho_{min} < 0.2$. In this way, we quickly created 294 families without performing property-driven optimization. 
Although efficient, this method also causes an imbalance in geometry, since several structure factors differ only by a coefficient and lead to overlapping families. For example, the equations for space groups No. 195 and 196 listed in~\cite{ITC2010crystaltables} are related as $A_{195,(hkl)} = 4 \cdot A_{196,(hkl)}$, and therefore generate the same structures. Next, we demonstrate the prowess of METASET in systematically removing such overlaps when selecting diverse subsets.

\subsection{Diverse 3D Families and Comparison of Shape Similarity Metrics}\label{sec:3ddpp}
While applying METASET to discover unique isosurface families, we also test the impact of the two proposed 3D shape similarity metrics (Sec.~\ref{sec:3dsim_ae}): the Hausdorff distance and the cosine similarity between deep learning-based embeddings.
As the families are comprised of a range of densities and thus shapes and properties, we need to capture the similarities of individual unit cells while assessing the similarities between families. Therefore, we generate 100 samples from each family covering the feasible range identified in the previous section, giving 29,400 unit cells total. Each unit cell is represented as a 4096-dimensional point cloud by first converting its level-set field into a triangle mesh~\cite{Vogiatzis2017mesh}, and then sampling on the triangular faces~\cite{sample_mesh}. We also remove any small disconnected features during post-processing, and find the homogenized elastic tensors of each unit cell using a code modified from~\cite{Dong2018homogenization}.

To quantify the similarity between two families, we assume each family is a collection of points, where each point corresponds to a unit cell. This reduces the problem of finding similarity between two families to one between two point sets using the Hausdorff distance (Eq.~\ref{eq:hausdorff}).
We calculate the similarity between families $C$ and $D$ in two steps: first using one of the 3D metrics to calculate the distance between individual unit cells $c \in C$ and $d \in D$, and then substituting this into the Hausdorff distance to obtain the \textit{inter-familial} distance, $h(C,D)$. 
Intuitively, this means that the shape similarity between two families is the maximum of the similarities between closest-in-shape pairs of unit cells. 
In property space, the similarity between families is related to the maximum of the pairwise Euclidean distances between each unit cell's properties. 
Therefore, rather than simply averaging the features of each family, the inter-familial similarities also consider the diversity of individuals within each family.
In short, we apply METASET to our 3D dataset using two approaches to measure shape similarity:
\begin{itemize}
\setlength{\itemsep}{0pt}
    \item H-H: Hausdorff distance between unit cells, followed by Hausdorff distance between families
    \item E-H: embedded cosine similarity between unit cells, followed by Hausdorff distance between families.
\end{itemize}

\begin{figure}[htb]
    \centering
    \begin{subfigure}{0.8\columnwidth}
        \includegraphics[width=1\columnwidth]{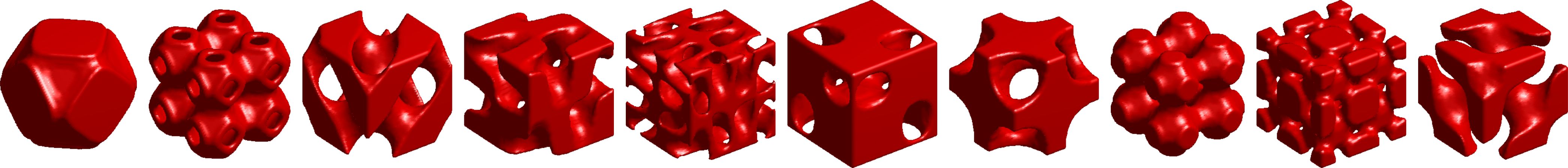}
        \caption{Families diverse in property space ($w=0$)}
        \label{fig:3d_0}
    \end{subfigure}
    
    \begin{subfigure}{0.8\columnwidth}
        \includegraphics[width=1\columnwidth]{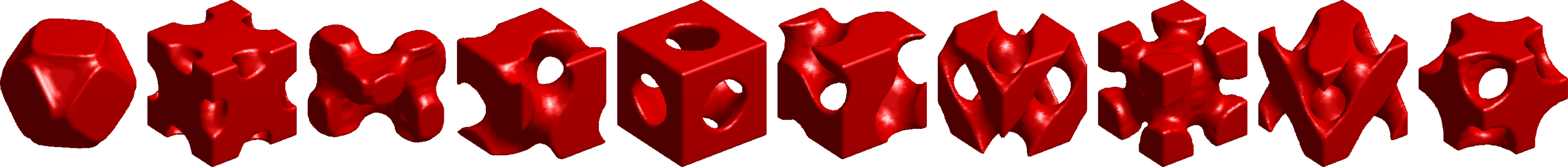}
        \caption{Families diverse in shape and property spaces ($w=0.5$)}
        \label{fig:3d_05}
    \end{subfigure}
    
    \begin{subfigure}{0.8\columnwidth}
        \includegraphics[width=1\columnwidth]{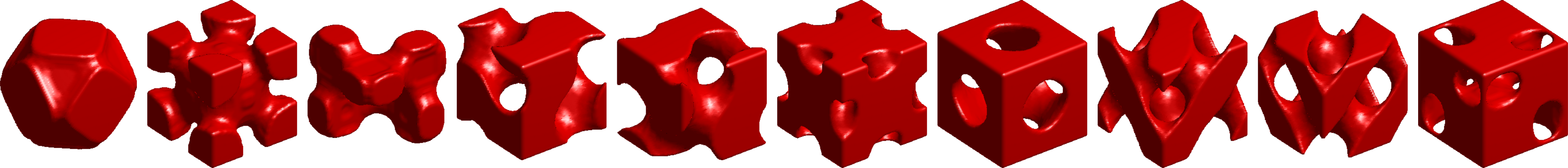}
        \caption{Families diverse in shape space ($w=1$)}
        \label{fig:3d_10}
    \end{subfigure}
    
    \caption{Examples of subsets of 3D isosurface families selected by METASET using the H-H shape metric.}
    \label{fig:3d_sets}
\end{figure}

\begin{figure}[htb]
    \centering
    \includegraphics[width=0.8\columnwidth]{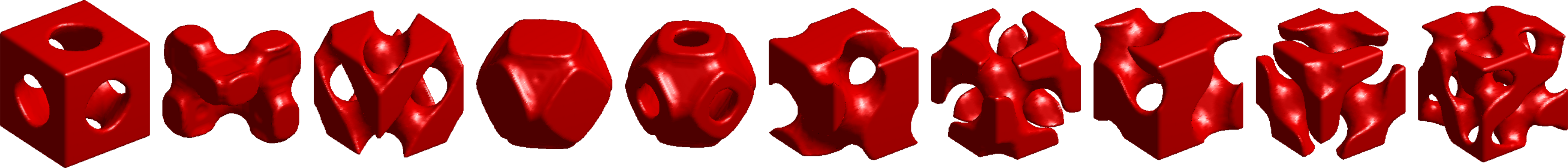}
    \caption{Shape diverse subset ($w=1$) selected by METASET using the embedding-based E-H shape metric.}
    \label{fig:3d_sets_cosine}
\end{figure}

Utilizing both of our shape similarity metrics, along with the property metric, we find diverse subsets of $10$ isosurface families. In addition, we vary the joint diversity weight $w$ between $0$ and $1$ (Eq.~\ref{eq:jointkernel}).
Some example subsets of diverse families are shown in Figs.~\ref{fig:3d_sets} and~\ref{fig:3d_sets_cosine}, where the median sample from each family are pictured. Like the 2D diverse subsets (Sec.~\ref{sec:2ddpp}), the property-only and shape-only sets (Figs.~\ref{fig:3d_0} and \ref{fig:3d_10}, respectively) share very few of the same families. 
Intriguingly, the shape diverse sets obtained from either metric contain families generated from the same space group and $(hkl)$, but different level-set forms (Eq.~\ref{eq:SF_analytical}). For example, with the H-H approach, the fourth and fifth items in Fig.~\ref{fig:3d_10} have the equations $A_{213,(011)}\geq t$ and $A_{213,(011)}\leq t$. The same families appear with the E-H approach as the eighth and sixth items. One could think of these as completely different shapes with almost no overlaps, which is further validation of the shape diversity chosen by METASET. 

\begin{table}[ht]
\centering
\caption{Shape diversity scores of subsets of 10 isosurface families, evaluated using either the Hausdorff (H-H) or embedded (E-H) shape metrics. The first two columns are for shape diverse subsets selected by METASET; the last shows the maximum of 10,000 random sets. The highest scores of each row in \textbf{bold} indicate that METASET always maximizes the diversity score with respect to the metric used during selection.}
\label{tab:3d_max}
\begin{tabular}{@{}llll@{}}
\toprule 
 & METASET (H-H) & METASET (E-H) & Random Sampling \\ \midrule
Score (H-H) & \textbf{1.0554E-04} & 6.3690E-05 & 2.8504E-05 \\
Score (E-H) & 1.6250E-13 & \textbf{6.4271E-12} & 5.4262E-14 \\
\bottomrule
\end{tabular}
\end{table}

Comparing the subsets obtained via either similarity metric (Figs.~\ref{fig:3d_10} and~\ref{fig:3d_sets_cosine}), we observe that 6 out of 10 shape diverse families overlap, indicating that the choice of metric does not drastically impact diversification. 
This is supported by a correlation coefficient of 0.836 between the H-H and E-H shape similarity kernels, $L_S$, of the shape diverse families.
Additionally, we cross-examine these results by applying the E-H approach to score ($det(L_S)$) the shape diverse subset chosen using H-H, and vice versa. As a baseline, we also randomly sample 10,000 sets of 10 families without METASET and measure their diversity with respect to each metric. The results are reported in Table~\ref{tab:3d_max}, where the greatest (most diverse) scores across each row reveal that the greedy algorithm will maximize the diversity score regardless of the similarity metric employed. Moreover, the subsets chosen by METASET have higher diversity than the random ones no matter which metric is used to evaluate the score. 

\begin{figure}[htb]
    \centering
   \includegraphics[width=0.55\columnwidth]{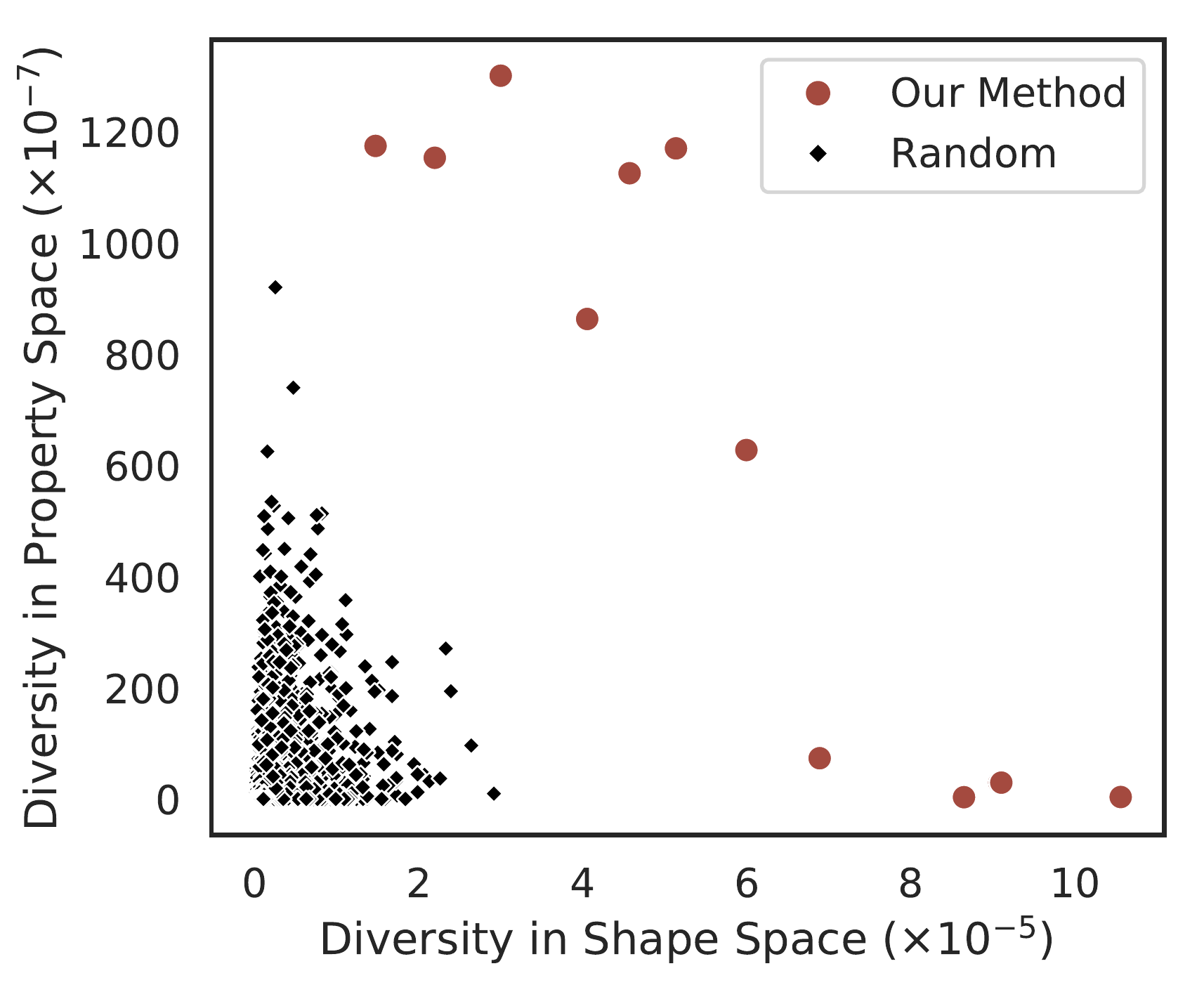}
    \caption{Trade-off between diversity in property vs. shape (using the H-H approach) spaces. The minimum diversity in shape space for optimized sets has a diversity score greater than 99.74\% of random samples.}
    \label{fig:3d_dpp_tradeoff}
\end{figure}

The high diversity of our subsets can also be seen in Fig.~\ref{fig:3d_dpp_tradeoff}, where their property and shape scores using H-H are plotted against those of the 10,000 random subsets.
Here, 99.74\% of the random sets (which are representative of the distribution of pairwise similarity values for our dataset) still fall short of the optimized subset with the lowest shape diversity score. This is compelling evidence that 1) the original dataset was severely imbalanced, and 2) METASET is able to combat such bias and select more diverse subsets. 

Fig.~\ref{fig:3d_dpp_tradeoff} additionally visualizes the trade-off between diversity in the shape and property spaces. Although our greedy algorithm maximizes the joint diversity score, the independent shape and property scores illustrate that, in general, the diversity in one space drops as we select sets that are more diverse in the other. 
This trade-off might raise a question as to whether a set of families that are quite diverse in property space can have low diversity in shapes, even though similar shapes are expected to possess similar properties. Our previous observation emerges as an answer: the sets of families with higher diversity in property space and seemingly ``low'' diversity in shape space actually have larger shape scores than the majority of the random sets. Therefore, the highest diversity in property space is achieved by a set of families which are also very diverse in shape.

Finally, we note that our diversified sets include isosurface families beyond the common TPMS used in existing metamaterials design, such as the Primitive, Gyroid and Diamond (see~\cite{Maskery2018tpms,Li2019tpms}).  
We provide the data of the METASET results publicly so that our diverse families can be employed by any designer in their work as well.
For example, these can be directly utilized in existing functionally-graded design methods such as~\cite{Li2019tpms}. Data-driven design with diverse isosurface families will be investigated in future works.

\section{Discussion}\label{sec:discussion}
Although we illustrated the benefits of METASET with several case studies, there are nevertheless some topics worthy of examining in the future. From our design of 2D aperiodic structures, we saw that shape diverse subsets may increase the chance to find compatible neighboring unit cells, while property diverse sets might enhance problems that require a wider range of target properties at the cost of connectivity. This dependence on shape vs. property diversity extends to ML tasks in the data-driven design framework (Fig.~\ref{fig:flow}) as well. To train property prediction models, one may need a property diverse dataset, while for a deep generative model that learns geometric features, a shape diverse set might be more appropriate. Along these lines, it would be interesting to further validate the improved performance of design and ML tasks using our subsets of diverse 3D unit cell families in a future work.

In the 2D examples, we also observed that smaller subsets led to designs with performance closer to the targets; in fact, we found using METASET that only 20 unit cells were enough to form a diverse subset. In most cases, the benefits of reducing the search space, model training time, or storage requirement of the dataset could outweigh any loss of data. However, certain applications such as ML may need large datasets. 
A key benefit of using a METASET, even for large subset sizes, is that it reduces bias by rank ordering all items in the dataset. The items with the highest redundancy in shape or property (like duplicates) are pushed toward the end of the rank-ordered list, so that ML algorithms trained on any subset will be less biased.
While it is not difficult to increase the size of the set, determining how much data is enough is more challenging since this too is contingent on the application and any limits on computational cost.
The effectiveness of size and diversity on specific tasks in metamaterials design is an important question for future studies.
Fortunately, the ease at which a subset's size as well as the weight of shape and property diversity can be explored is yet another advantage of METASET.

Lastly, we remark that the capability of METASET depends on the choice of similarity metrics as well as the definition of the joint similarity kernel, both of which are avenues of further research. Our 3D study demonstrated that METASET will maximize diversity regardless of the metric adopted. Although the results indicated that different shape similarity metrics can be highly correlated and slightly change the diverse subsets, there are a wealth of other choices that may provide different results. Extending METASET to more complex properties, like dynamic ones, may necessitate new metrics. For the joint DPP kernel, we chose a simple weighted sum to join the shape and property matrices, thereby casting the greedy selection as a multi-objective problem. We found in Sec.~\ref{sec:2ddpp} that this was a valid assumption, but other methods to combine kernels while preserving submodularity are also possible. However, swapping these to best suit the application is easily done since the input of the DPPs-based greedy algorithm in METASET is a positive semi-definite similarity kernel that can be obtained from any appropriate metric or definition.

\section{Conclusion}\label{sec:conclusion}
In this paper, we proposed a methodology, METASET, that incorporates joint diversity in the shape and property spaces into data selection to improve the downstream tasks in data-driven design. As an enhancement to any existing data-driven framework, METASET is efficient and flexible, allowing the emphasis on either shape or property to be easily traded by measuring and maximizing the joint diversity of subsets through a weighted DPP similarity kernel. To calculate this kernel matrix, we introduced similarity metrics that cater specifically to 2D and 3D metamaterials.

By way of our 2D aperiodic metamaterial design examples, we demonstrated that small yet diverse subsets of unit cells can boost the scalability of search algorithms while leading to designs with greater performance and enhanced boundary compatibility. This revelation shakes a common belief in the field of data-driven mechanical metamaterials design that a larger and denser dataset is required to design well-connected structures while still meeting the target behavior. To our knowledge, this is the first time that such a result has been studied and presented.

In our 3D case study, we not only proposed a new method to generate triply periodic isosurface unit cells using crystallographic structure factors, but also verified that METASET can effectively discover unique unit cell families in order to build diverse, unbiased and economical datasets for design regardless of the shape similarity metric employed. Different from well-known TPMS unit cells, our dataset of families are optimized for shape and property diversity rather than arbitrarily chosen. In future works, we will explore the use these diverse families for data-driven metamaterials design and ML.

 Although this paper focused on showcasing METASET through the design of mechanical metamaterials, the methods we proposed are broadly applicable to other metamaterial domains, or indeed any other design problems that need to balance design space against some performance or quality space. In design ideation, our method can be used to select ideas that are functionally different from each other while achieving different performance goals. It can also be integrated with existing multi-objective optimization algorithms as a niching method. To contribute to the growth and capability of data-driven metamaterials design methods and other fields, we have shared diversified subsets of 2D and 3D unit cells, as well as the corresponding equations of isosurface families. These unit cells can be directly plugged into the application of any metamaterials designer.

\section*{Acknowledgements}
We are grateful for support from the National Science Foundation (NSF) CSSI program (Grant No.~OAC-1835782). Yu-Chin Chan thanks the NSF Graduate Research Fellowship (Grant No.~DGE-1842165). Liwei Wang acknowledges support from the Zhiyuan Honors Program for Graduate Students of Shanghai Jiao Tong University for his predoctoral visiting study at Northwestern University.

\section*{Data Availability}
The diverse datasets autonomously selected by the METASET method can be found at \url{https://github.com/lychan110/metaset}.

\bibliographystyle{unsrtnat}
\bibliography{ms}

\end{document}